\documentclass[
superscriptaddress,
nofootinbib,
amsmath,amssymb,
aps,
pre,
longbibliography,
showkeys
]{revtex4-1}
\usepackage[left=2.5cm,top=3cm,right=2.5cm,bottom=3cm,bindingoffset=0.5cm]{geometry}
\usepackage{graphicx}
\usepackage{hyperref}
\usepackage[bitstream-charter, greekfamily=default]{mathdesign}
\usepackage{lipsum}

\usepackage{xcolor}

\newcommand{\de}{\mathrm{d}}

\begin{document}
\begin{center}
   \Large \textbf{Electronic states of a disordered $2D$ quasiperiodic tiling: from critical states to Anderson localization}\vspace{0.1cm}
   \hrule
\end{center}

\author{Anuradha Jagannathan}
\affiliation{
Laboratoire de Physique des Solides, B\^at.510, Universit\'{e} Paris-Saclay, 91405 Orsay, France
}
\author {Marco Tarzia}
\affiliation{ Laboratoire Th\'eorique de la Mati\`ere Condens\'ee, Sorbonne Universit\'e, 75252 Paris C\'edex 05, France}
\affiliation{Institut  Universitaire  de  France,  1  rue  Descartes,  75231  Paris  Cedex  05,  France}

\date{\today}

    \begin{abstract}
We consider critical eigenstates in a two dimensional quasicrystal and their evolution as a function of disorder. By exact diagonalization of finite size systems we show that the evolution of properties of a typical wave-function is non-monotonic. That is, disorder leads to states delocalizing, until a certain crossover disorder strength is attained, after which they start to localize. Although this non-monotonic behavior is only present in finite-size systems and vanishes in the thermodynamic limit, the crossover disorder strength decreases logarithmically slowly with system size, and is quite large even for very large approximants. The non-monotonic evolution of spatial properties of eigenstates can be observed in the anomalous dimensions of the wave-function amplitudes, in their multifractal spectra, and in their dynamical properties. We compute the two-point correlation functions of wave-function amplitudes and show that these follow power laws in distance and energy, consistent with the idea that wave-functions retain their multifractal structure on a scale which depends on disorder strength. Dynamical properties are studied as a function of disorder. We find that the diffusion exponents do not reflect the non-monotonic wave-function evolution. Instead, they are essentially independent of disorder until disorder increases beyond the crossover value, after which they decrease rapidly, until the strong localization regime is reached. The differences between our results and earlier studies on geometrically disordered ``phason-flip'' models lead us to propose that the two models are in different universality classes. We conclude by discussing some implications of our results for transport and a proposal for a Mott hopping mechanism between power law localized wave-functions, in moderately disordered quasicrystals.
    \end{abstract}
    
\maketitle

\section{Introduction}
\begin{figure}
\includegraphics[width=0.6\textwidth]{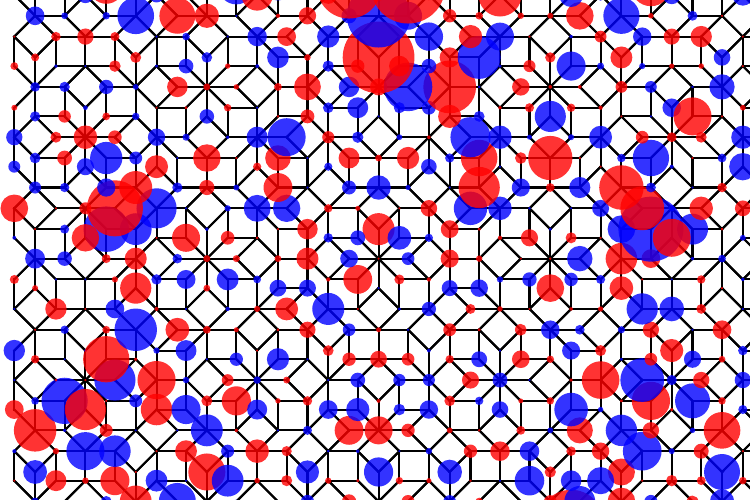} 
\caption{A patch of the Ammann-Beenker tiling illustrating a critical wave-function (for model Eq.~\ref{eq:hamiltonian} without disorder and $E=0.24t$). The magnitude and sign of the wave-function amplitude on each site are represented by the circle radius and color respectively.}
\label{fig:tiling}
\end{figure}

The electronic properties of quasicrystals have continued to pose fundamental problems ever since the discovery of these materials by Schechtman et al \cite{schechtman}. The question of what constitutes ``intrinsic'' behavior of quasicrystals, and how it is affected by disorder is not yet clearly understood. The reasons for this deficit of understanding lie in the fact that the complexity of real materials -- both chemical and structural -- render numerical computations difficult, and that analytical tools lack for even the simplest two or three dimensional cases.  
Many numerical studies have been carried out for simplified tight-binding models on quasiperiodic tilings. These indicate that single particle electronic states are, typically, multifractal for undisordered tilings (see review in \cite{baakegrimm}). The generalized dimensions can in fact be exactly computed for the ground states of two well-known tilings \cite{mace2017}. However, it is not understood what happens to such multifractal states in the presence of disorder. The nature of electronic states in real quasicrystals, where a certain amount of disorder is inevitably present, has remained an open question. While the transition to Anderson localization has been studied in detail for a 1D quasicrystal, the Fibonacci chain \cite{disorderFC1,disorderFC2}, the effects of disorder on eigenstates have not been analyzed in detail for 2D and 3D quasicrystals.  
This paper fills the gap by presenting a comprehensive numerical study of the effects of disorder on eigenstates of an eight-fold symmetric 2D tiling, the Ammann-Beenker (AB) tiling \cite{grunbaum}. A patch of this tiling, composed of squares and $45^\circ$ rhombuses is shown in Fig.~\ref{fig:tiling}, along with a typical critical state, represented by colored circles whose radii are proportional to the amplitude on each site. We will address here in detail the question of how such multifractal or critical states evolve as a function of disorder into strongly Anderson localized states. 

In the study of disordered quasiperiodic tilings the extensive literature on a variety of disordered systems provides a valuable guide. Random matrix models \cite{mehta} are important reference systems which predict certain universal properties.  In particular, as we will describe, the Gaussian orthogonal ensemble (GOE) which has the symmetry of the Hamiltonians we will consider here describes some spectral properties of tilings for both the pure case and for weakly disordered tilings of finite size. The best known reference system for condensed matter is the Anderson model for localization in random lattices \cite{anderson}. In the Anderson model, it is well-known that a metal-insulator transition occurs for three dimensional lattices, in contrast to 2D or 1D lattices, where any disorder, however small, suffices to localize all of the eigenstates~\cite{scaling}. These statements hold in an infinite system. However, if one considers a finite system of linear dimension $L$, one can distinguish between two regimes of disorder  -- a weak disorder regime, where states appear to be extended on the scale of the system size, and a strong disorder regime, where the states are clearly localized. The crossover occurs for a disorder strength $W_\star$ which corresponds to the localization length $\xi_{\rm loc} (W_\star) \sim L$. Since in 2D the localization length diverges exponentially as $W \to 0$, the apparent metallic-like regime at weak disorder extends up to very large system sizes.

We show here that, within the weak disorder regime, most states display a non-monotonic behavior as a function of the disorder strength $W$. That is, states tend at first to delocalize for very small $W$. Localization occurs only when the disorder exceeds a crossover value $W_\star$, which depends on the energy and system size. The non-monotonic evolution of the localization
properties only exists for finite systems, since the value of the cross-over disorder strength vanishes in the thermodynamic limit. However the characteristic disorder $W_\star$ decreases logarithmically with the system size and hence the non-monotonicity is observed also for very large approximants. The non-monotonic behavior can be seen by plotting the inverse participation ratio (IPR) as a function of disorder, but also in the $f(\alpha)$ curve, which describes multifractal properties of wave-functions, and in the generalized dimensions of the wave functions $D^\psi_q$. 
This non-monotonicity is particularly pronounced for the states around zero energy, in the middle of the spectrum. This is due to the fact that there is a zero-width band of localized states at $E=0$ in this model, analogous to the $E=0$ states present in the well-known Penrose rhombus tiling ~\cite{kohmotosutherland,arai}.  All of these degenerate states are mixed as soon as disorder is introduced, and they delocalize rapidly for $W\neq 0$. In contrast, the states lying near band edges and the main pseudogap behave differently. They are the fastest to localize and, upto numerical accuracy, their evolution under disorder is monotonic.

To complete the description of electronic properties in the disordered quasicrystal, we have carried out detailed studies of
two-point spatial correlation functions $C(r,\omega;E)$. These describe correlations of the amplitudes of two eigenstates of energies close to $E$, having an energy separation $\omega$, as a function of spatial distance $r$. We obtain these correlation functions for increasing values of disorder, and show that for a large range of disorder strength across the crossover regime, correlations follow power laws over a broad range of distances. In the crossover regime, the wave function correlations are seen furthermore to be enhanced by disorder.

Dynamical properties are then studied as a function of disorder strength. 
We consider the spreading of wavepackets which are initially localized on clusters composed of a central site of connectivity $k$ ($3 \le k \le 8$ for the AB tiling) and all of its nearest neighbors. The initial wavepacket energy is varied by a parameter $\theta$, which controls the initial wave function amplitudes on the center of the clusters and on the external nodes.
We study the probability of return to the origin (cluster center) of such wavepackets released at time $t=0$, $P(t)$, which is closely related to the Fourier transform of the correlation function $C(0,\omega;E)$ (computed for amplitudes at the same spatial point, $r=0$), and is found likewise to exhibit different regimes of behaviors as a function of $t$. After an initial fast decay for very short times, $P(t)$ shows a power law decay in a wide range of $t$ and ultimately reaches a constant value at very long times. The onset of this power law regime moves to shorter times with increasing disorder strength. The addition of quenched disorder has therefore the effect of promoting the spreading of the wave packet in the crossover regime. This effect is particularly strong for small $k$. We next computed the mean square distance, given by the average over all initial positions and over disorder realizations of $\langle r^2(t)\rangle \sim t^{2\beta}$. We measured the effective diffusion exponents $\beta$ (which depend on $k$ and $\theta$) as a function of disorder strength. When disorder is turned on $\beta$ as a function of $W$ appears initially to remain constant. It then drops quickly to zero when the localization regime is reached for sufficiently strong $W \gtrsim W_\star$. 

A number of previous studies have investigated the effects of geometrically disordered tilings. These models modify the quasiperiodic structure by making local permutations of the tiles at random locations, termed ``phason-flips''. Such models were studied both in 2D \cite{benza92,piechon95,schwabe,yamamoto,vekilov99} and 3D \cite{vekilov00}. These studies of spectral and quantum diffusion properties concluded that phason disorder tends to delocalize the electronic states and promote quantum diffusion. The conductivity in finite samples of the 2D Penrose tiling was computed using Kubo formalism, and found to increase with a type of phason disorder \cite{fujiwara1993}. 
Grimm and Roemer carried out a recent study of spectral properties and found that both the pure and the phason-disordered tilings obey GOE statistics, implying absence of localization~\cite{grimmroemer}. The spectral properties of phason-disordered models agree qualitatively with those of our model, in the weak disorder limit. However the diffusion exponent $\beta$ in our model does not show an increase for weak disorder, in contrast to the results reported in Ref.~\cite{benza92}. This may imply that the effects of random on-site energies on quantum diffusion and transport are different from those of phason disorder. For phason disordered models no localized phase has been observed, contrarily to the onsite disorder model. It is an open question as to whether one can achieve an Anderson localized phase by phason disorder alone.

As stated at the outset, one of the motivations for this study is to understand electronic properties of quasicrystals, in particular, transport. Experimental measurements of electrical conductivity of quasicrystals show that they are strikingly poor conductors compared to their constituent metals. One of the contributing factors is that many quasicrystals display a strongly reduced density of states (DOS) at the Fermi level compared to the parent materials -- the ``pseudogap''. The second factor has to do with the nature of the eigenstates, and the resulting anomalous diffusion in quasicrystals.
Weak localization theory and electron-electron interactions theory have been invoked to explain the roughly linear increase of conductivity with temperature observed in very high structural quality icosahedral quasicrystals such as i-AlCuFe \cite{ahlgren,chernikov,rodmar}. In contrast, however, the low temperature transport in i-AlCuRe samples obtained by fast-quench has given rise to much debate. On the one hand, their DOS is very similar to that of samples obtained by slow cooling \cite{nayak,sarkar}. Yet, there are huge differences in transport -- and the former have insulating behavior, as confirmed by recent transport measurement which reported Mott variable range hopping (VRH) between localized states at low temperatures. The answer evidently lies in the disorder strength and differences of the resulting electronic states in the two sets of samples. Based on our study, we propose that for the more disordered samples it may be possible to observe power law conductivity due to hopping between disorder-modified critical states. In addition there could occur a crossover to the usual Mott exponential dependence for localized states at very low temperature $T_{co}$. 

The paper is organized as follows: in Sec.~\ref{sec:dos} the model is introduced, and the DOS is shown for different values of disorder. Sec.~\ref{sec:ipr} presents the disorder dependence of the inverse participation-ratio of the eigenstates, which show the non-monotonic behavior. In Sec.~\ref{sec:multi} we focus on the anomalous dimensions of wave-functions' amplitudes and on the multifractal analysis of the singularity spectrum. Sec.~\ref{sec:corr} discusses two-points correlation functions. Sec.~\ref{sec:dyn} considers dynamical properties: the probability of return to the origin and  the power law spreading of wave packets. Sec.~\ref{sec:vrh} presents a discussion of transport based on the results. Sec.~\ref{sec:conclusions} ends with a discussion and conclusions.

\section{The model and its density of states} \label{sec:dos}

The tight-binding model we study is described by the Hamiltonian
\begin{eqnarray}
H = \sum_{ i} \epsilon_{i} c^\dag_ic_i + \sum_{\langle i,j\rangle} (t_{ij} c^\dag_ic_j + h.c.)
\label{eq:hamiltonian}
\end{eqnarray}
where $i=1,...,N$ are site indices, and ${\langle i,j\rangle}$ denotes pairs of sites that are linked by an edge (see Fig.~\ref{fig:tiling}). Spin indices are not written as they play no role aside from introducing a factor 2. The systems considered are square approximants -- tilings which are periodic but resemble the infinite quasiperiodic tiling in their local geometry --  generated from a 4D hypercubic lattice by the cut-and-project method \cite{duneaumoss}. In addition to the eight-fold symmetry the tiling possesses a discrete scale invariance --  a tiling of edge length $a$ can be transformed into a tiling of edge lengths which are smaller by a scale factor $\lambda = 1+\sqrt{2}$ and vice-versa (called inflation/deflation transformations, see ~\cite{grunbaum}). In our computations, square approximants of total number of sites $N$ equal to 239, 1393, 8119 and 47321 were considered. Periodic boundary conditions were imposed in both directions.

In the pure limit, all the onsite energies are taken to be equal $\epsilon_i=\epsilon_0$, and all hopping amplitudes to be equal $t_{ij}=t$. Without loss of generality, the origin of the energy is chosen such that $\epsilon_0=0$, and energy units are chosen so that $t=1$. Since the tiling is bipartite (made from quadrilaterals) the energy spectrum of this model is symmetric. Thus all plots are shown henceforth for the positive half of the spectrum only. The scale invariance of the tiling has been used in renormalization group schemes for electronic states \cite{siremoss90} and in the calculation of ground state properties of an antiferromagnetic Heisenberg model  \cite{jagannathanPRL}.
Moreover, since the AB tiling can be generated from a parent 4-dimensional cubic lattice, the Hamiltonian Eq.~\eqref{eq:hamiltonian} can be related to a four dimensional Quantum Hall system \cite{zilberberg}, and inherits certain topological properties which will not however concern us here.

\begin{figure}[h]
\includegraphics[width=0.6\textwidth]{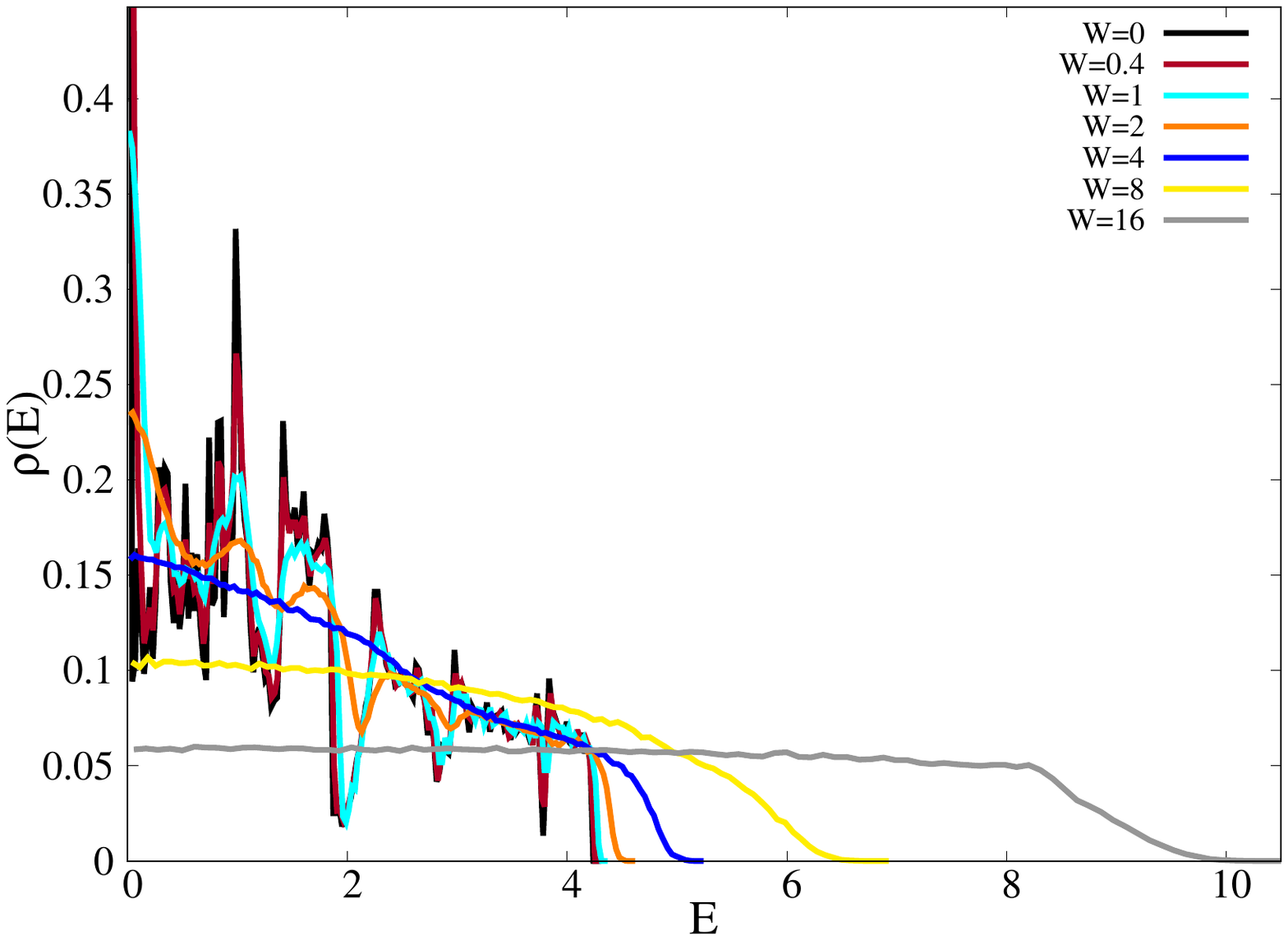} 
\caption{Density of states $\rho(E)$ for several values of the disorder and for $N=47321$. 
\label{fig:dos}}
\end{figure}

The DOS of the pure quasiperiodic tiling has a characteristic spiky shape (see Fig.~\ref{fig:dos} for $W=0$). The spiky local maxima and minima of the DOS are consequences of the symmetries of the Hamiltonian Eq.~\eqref{eq:hamiltonian}, due to the properties of the underlying structure such as scale invariance and repetitivity of environments.\footnote{The repetitivity property says that local configurations are guaranteed to repeat throughout the tiling with a maximal and a minimal allowed spacing, and generalizes the notion of strict translational invariance present in periodic crystals.}  Among the sharp minima or ``pseudogaps'', the most prominent pair is located near $E\approx\pm 1.95$. The pseudogaps are expected to be located at special values of the integrated density of states (IDOS) given by a gap labeling scheme \cite{kelput}, with the main gaps corresponding to IDOS values that approach the values $\lambda^{-2}$ and its symmetric $1-\lambda^{-2}$ as the system size increases. 

In the pure model, additionally, one finds a delta-function peak of the DOS exactly at $E=0$, corresponding to a macroscopic number of``confined'' states. Similar $E=0$ states can be found for other bipartite tilings including the well-known Penrose tiling ~\cite{kohmotosutherland,arai}.  In the AB tiling, the smallest such $E=0$ state is a small ring, with non-zero amplitudes on the 8 nearest neighbors of sites with a coordination number 8. The degeneracy of this type of confined state is given by $N\lambda^{-4}$ as $N\rightarrow \infty$. Other confined states can be similarly enumerated. The spatial features of the set of confined states (which can overlap) and their degeneracies are discussed in ~\cite{koga17,oktel,kamenev}.  The support of these confined states is a finite fraction of sites of the tiling. They are unstable with respect to most forms of disorder. 

Adding disorder -- by varying the parameters in~\eqref{eq:hamiltonian}, for example, or by modifying the geometry by random phason flips -- breaks the  symmetries of $H$, and singularities of the DOS are progressively smoothed out. Roughly speaking, disorder broadens the Bragg peaks of the structure, thus reducing quantum interference due to backscattering. 

A number of early studies considered the evolution of  spectral properties under phason flip disorder via the analysis of level statistics (see the review \cite{philmag2007}). Grimm and Roemer have \cite{grimmroemer} returned recently to the problem, performing a careful analysis of the r-value (ratio of the gaps of consecutive levels) statistics~\cite{pal_huse}. They reach the conclusion that states in both the pure and the phason-disordered tilings were described by the GOE ensemble. In other words, phason disorder did not induce localization of states (in which case one would have found not GOE but Poissonian statistics). From this, it seems clear that the phason disorder does not suffice to drive the system into the strong localization regime, at least, for the system sizes considered. 

In this work, instead, disorder is specifically introduced by assuming the onsite energies to be random variables taken from a box distribution of width $W$ (i.e. $\epsilon_i$ are independent and identically distributed random variables uniformly taken in the interval $[-W/2,W/2]$). We expect that the addition of quenched disorder also in the $t_{ij}$ yields the same results. Differently from phason disorder, out model~\eqref{eq:hamiltonian} allows one to tune the disorder from very weak disorder (reproducing effects seen for the phason-flip disordered model) out to arbitrarily high disorder, when all states are strongly localized. 

Results for the DOS of pure and disordered tilings obtained from exact diagonalizations of $H$ are presented in Fig.~\ref{fig:dos}. Ensemble averages are performed over several independent realizations of the disorder, using, here and throughout the rest of the paper, $M$ copies of the system with $M \approx 64$ for $N=47321$, $M \approx 256$ for $N=8119$, $M \approx 8192$ for $N=1393$, and $M \approx 2 \cdot 10^4$ for $N=239$.

As can be seen in Fig.~\ref{fig:dos}, at small $W$ the DOS is only slightly modified by the perturbation. However, when $W$ exceeds a characteristic scale (which for this size is about $W_\star \in [0.1 ,0.4]$) the DOS is strongly affected by the disorder and progressively smoothed out. For $W=16$, which corresponds to the strongly localized limit, it is easy to see that the DOS is essentially given by the distribution of the random energies (i.e. a uniform box distribution), plus the exponential so-called ``Lifshitz tails'' at the edges of the spectrum \cite{lifshitz} (which extend up to an energy equal to the spectral edge of the pure case, $|E| \approx 4.2 + W/2$).

\section{The Inverse participation ratio} \label{sec:ipr}
One measure of the degree of delocalization of the wave-functions is provided by the generalized inverse participation ratios $I_q$, defined as follows:
\begin{equation} \label{eq:Pq}
I_q(N,E) = \frac{ \left \langle \sum_{i=1}^N \sum_{r=1}^N |\psi_i(r)|^{2q} \, \delta \left (E_i - E \right) \right \rangle } { \left \langle \sum_{i=1}^N \delta \left (E_i - E \right) \right \rangle } \, ,
\end{equation}
where $|\psi_i (r)|^2$ is the value at position $r$ of the $i$-th normalized eigenstate of $H$. The asymptotic scaling behavior of the $I_q$'s allows one to distinguish between fully delocalized, fully localized, and critical states. In fact asymptotically $I_q(N,E)$ behaves as $I_q(N,E) \propto N^{-\tau_q}$, where the mass exponents $\tau_q$ is defined as:
\begin{equation} \label{eq:D}
\tau_q = \left \{
\begin{array}{ll}
q-1 & \textrm{for fully delocalized wave-functions} \, , \\
0 & \textrm{for fully localized wave-functions} \, , \\
D^\psi_q(q-1)~~~ & \textrm{for multifractal wave-functions} \, .
\end{array}
\right.
\end{equation}
The $D^\psi_q$'s are called ``generalized fractal dimensions''. In the following we will mainly focus on the so-called inverse participation ratio (IPR) $I_2$, but similar results are found for other moments of the wave-functions coefficients. $I_2$ is essentially the measure of the inverse volume occupied by an eigenstate: it is proportional to $1/N$ for fully delocalized plane-waves, it tends to a finite value (proportional to $\xi_{\rm loc}^{-D}$) for exponentially localized wave-functions, and scales as $N^{-D^\psi_2}$ with $0<D^\psi_2<1$ for critical states. 

\begin{figure}
\includegraphics[width=0.431\textwidth]{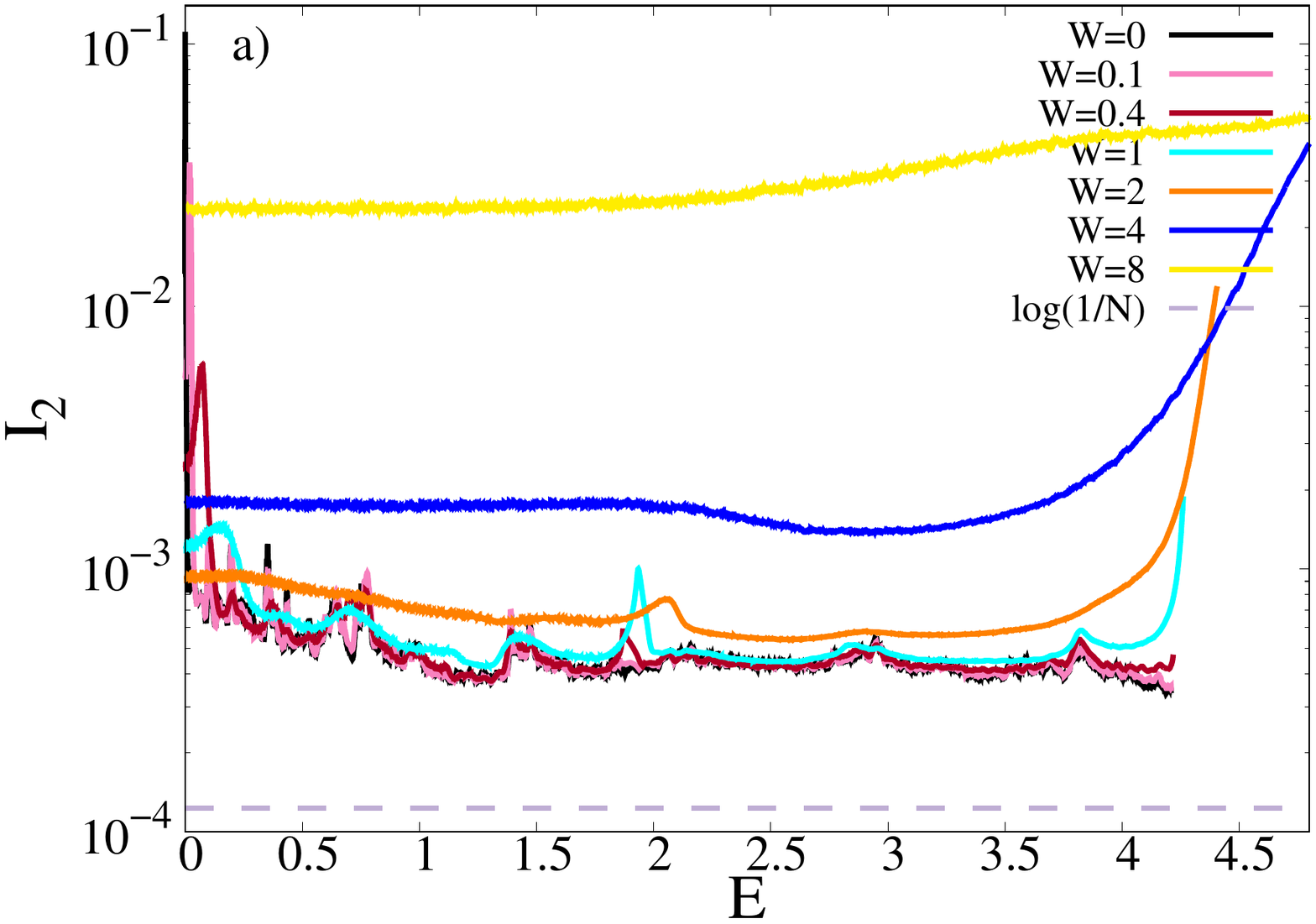} \hspace{0.1cm}
\includegraphics[width=0.431\textwidth]{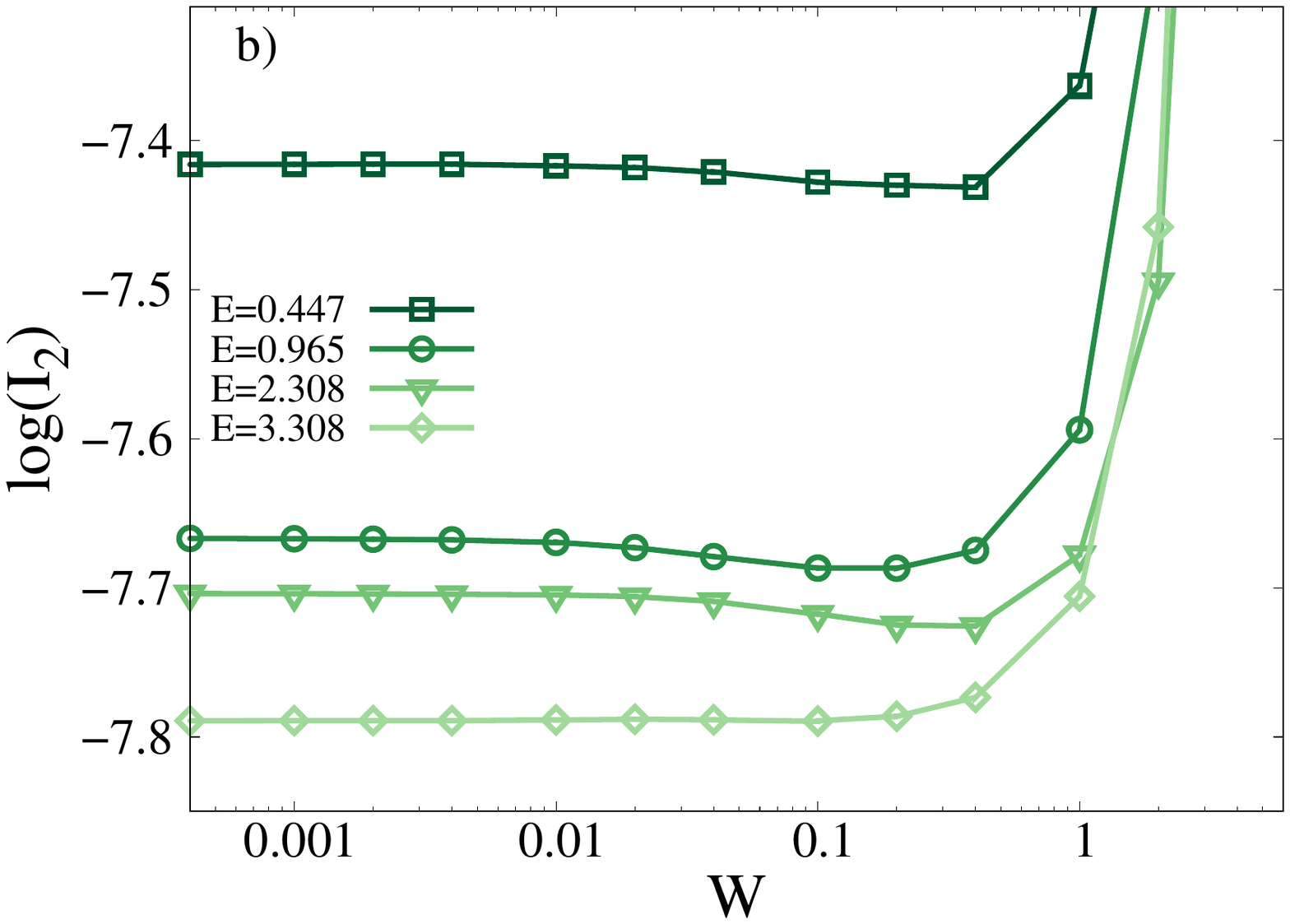}
\caption{a) $I_2$ as a function of $E$ for several values of the $W$ and for $N=8119$, across the whole spectrum. The IPR is averaged over several realizations of the quenched disorder and few eigenfunctions in a sliding energy window $[E-\delta,E+\delta]$, with $\delta =0.02$.  The horizontal dashed line corresponds to the minimal value of the IPR for fully extended plane-waves, $I_2 = 1/N$. b) $\log I_2$ as a function of $W$ for $N=8119$ and for several values of the energy across the energy band, showing the non-monotonicity.
\label{fig:ipr1}}
\end{figure}

The evolution of the IPR upon increasing the disorder for $N=8119$ are shown in Fig.~\ref{fig:ipr1}a) for all eigenstates across the whole spectrum.  The figure shows that in absence of disorder the critical eigenstates of the pure AB tiling are fairly extended (as the value of the IPR is only about $5$ times larger than the minimal value $1/N$ realized for fully extended plane waves). The states closer to the edge of the spectrum have typically a slightly smaller IPR compared to the states close to the middle of the band. The sharp peak of $I_2$ near $E=0$ is due to the confined states in the middle of the band which delocalize when $W\neq 0$. Many smaller peaks are seen in Fig.~\ref{fig:ipr1}a), notably the peak near the main pseudogap at $E\approx 1.95$. We will consider first the behavior of generic states in the band, and then that of states near the band center, and states near the main pseudogap and band edges which behave somewhat differently.

\begin{figure}
\includegraphics[width=0.331\textwidth]{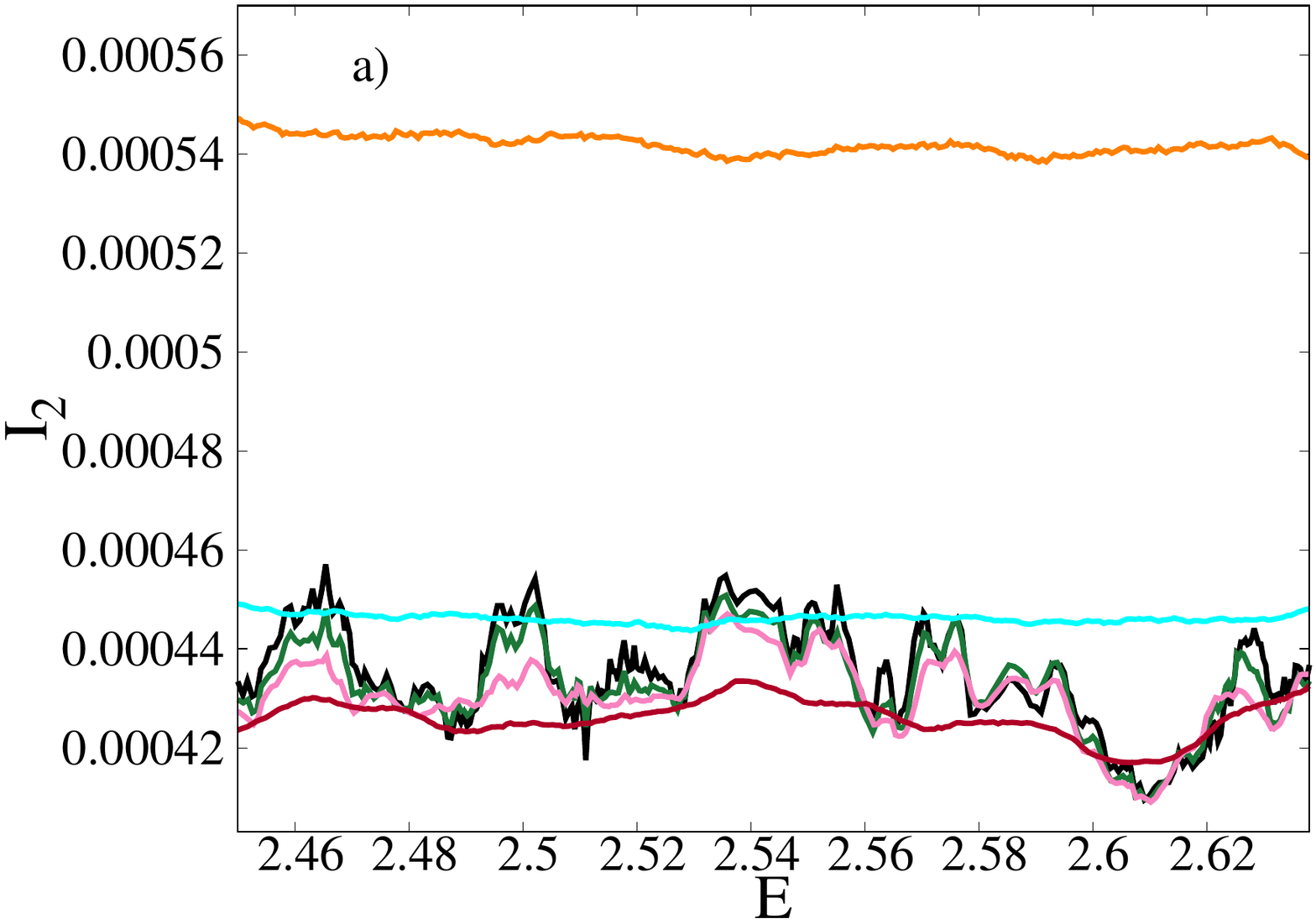} \hspace{-0.1cm} \includegraphics[width=0.331\textwidth]{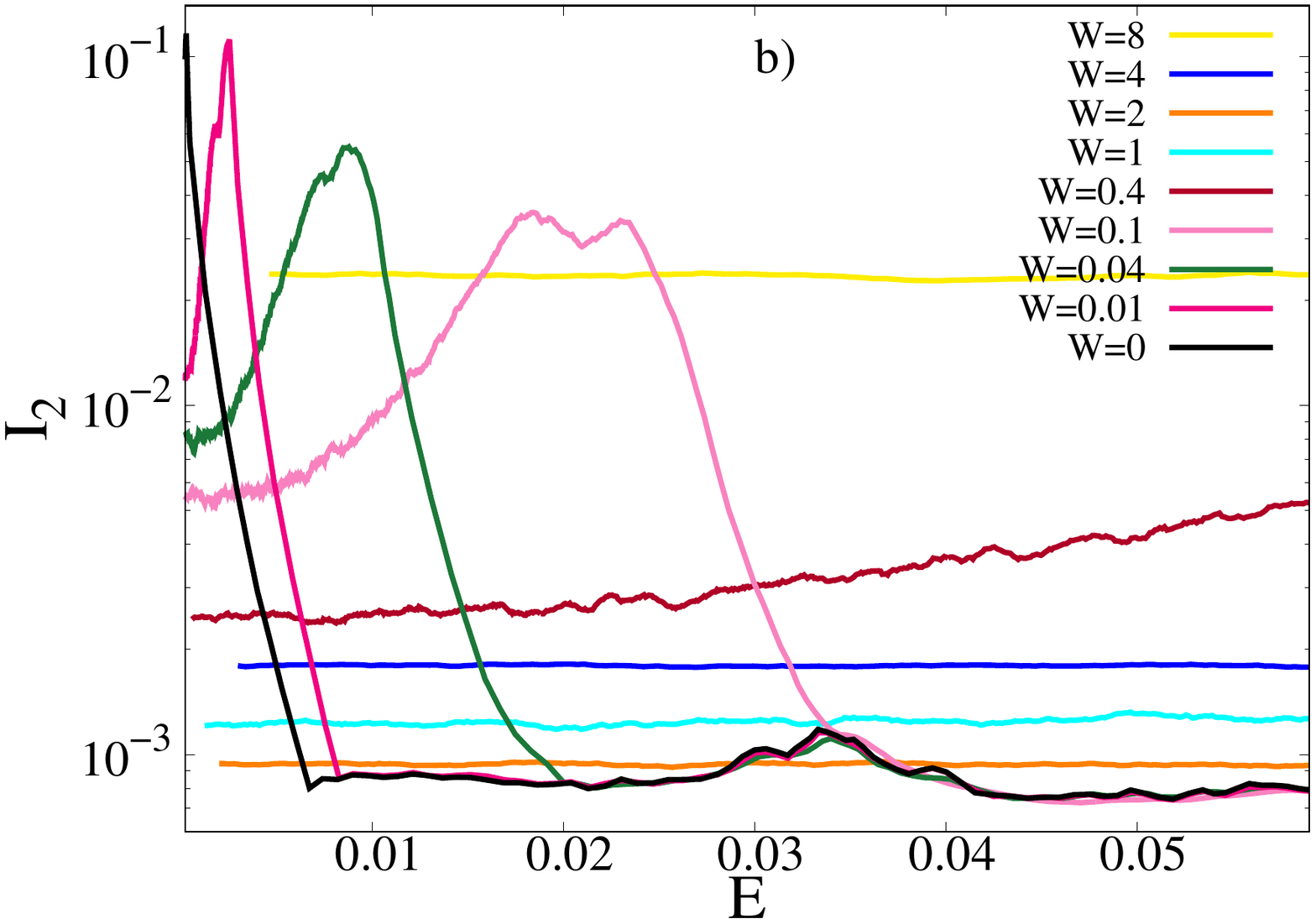} \hspace{-0.1cm} \includegraphics[width=0.331\textwidth]{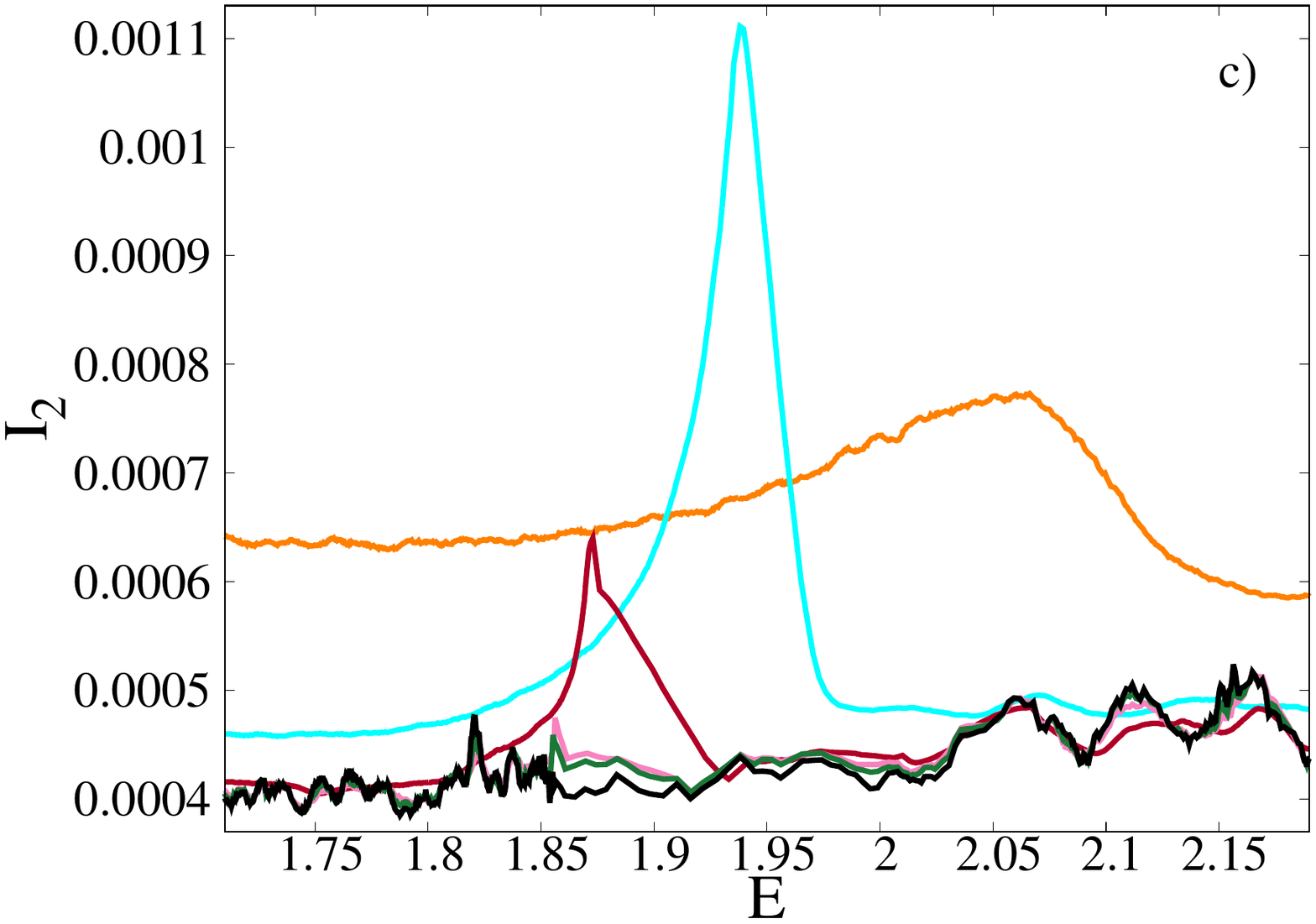} 
\caption{Plots of $I_2$ versus $E$ for several values of the $W$ in three selected regions of the spectrum, illustrating different behaviors (sample size $N=8119$). 
a) Zoom into the interval $E \in [2.45,2.63]$ showing the non-monotonic behavior characterizing the majority of states in the bulk of the spectrum, where the minimum of $I_2$ is attained either for $W=0.4$ (red) -- for the majority of the states -- or for $W=0.1$ (pink) -- for the states in the interval $E \in [2.6,2.62]$. The non-monotonicity is stronger for the states that have a higher $I_2$ in the pure limit. In this interval the only states for which the non-monotonic behavior is absent are at $E \sim 2.51$, for which the minimum of $I_2$ is obtained for $W=0$ (black). b) Zoom into the region $E \approx 0$ showing a stronger dependence on $W$. The $I_2$ of the localized states at zero energy first decreases rapidly to the low values found elsewhere in the spectrum, before increasing. Note also how the peak of the IPR of localized states is shifted to higher energies as $W$ is increased. c) Zoom into the region $E \approx 1.95$ showing that in the vicinity of the pseudogap, $I_2$ has no discernible initial dip (i.e., the lowest curve is the black one, corresponding to $W=0$) but seems to increase monotonically with $W$. The other characteristic of the states close to the pseudogap is the development of a sharp peak of $I_2$ as the disorder is increased above $W \sim 0.4$, signaling their strong sensitivity to the random perturbation.
\label{fig:ipr3}}
\end{figure}

{\it{Generic states in the bulk of the band:}} In Fig.~\ref{fig:ipr3}a) we shows a close-up of the IPR data in the energy interval $E \in [2.45,2.63]$. A close inspection of the curves indicates that  $I_2$ has a non-monotonic behavior for the majority of the states: it decreases initially and then starts to increase after reaching a minimum roughly in the interval $W_\star \in [0.07, 0.28]$. Such non-monotonicity is also highlighted in Fig.~\ref{fig:ipr1}b), where we plot $I_2$ as a function of the disorder for four selected values of the energy across the energy band, showing that $I_2$ first slightly decreases at small $W$ and then increases sharply at larger $W$, resulting in a minimum at the characteristic scale $W_\star$. 

\begin{figure}
\includegraphics[width=0.431\textwidth]{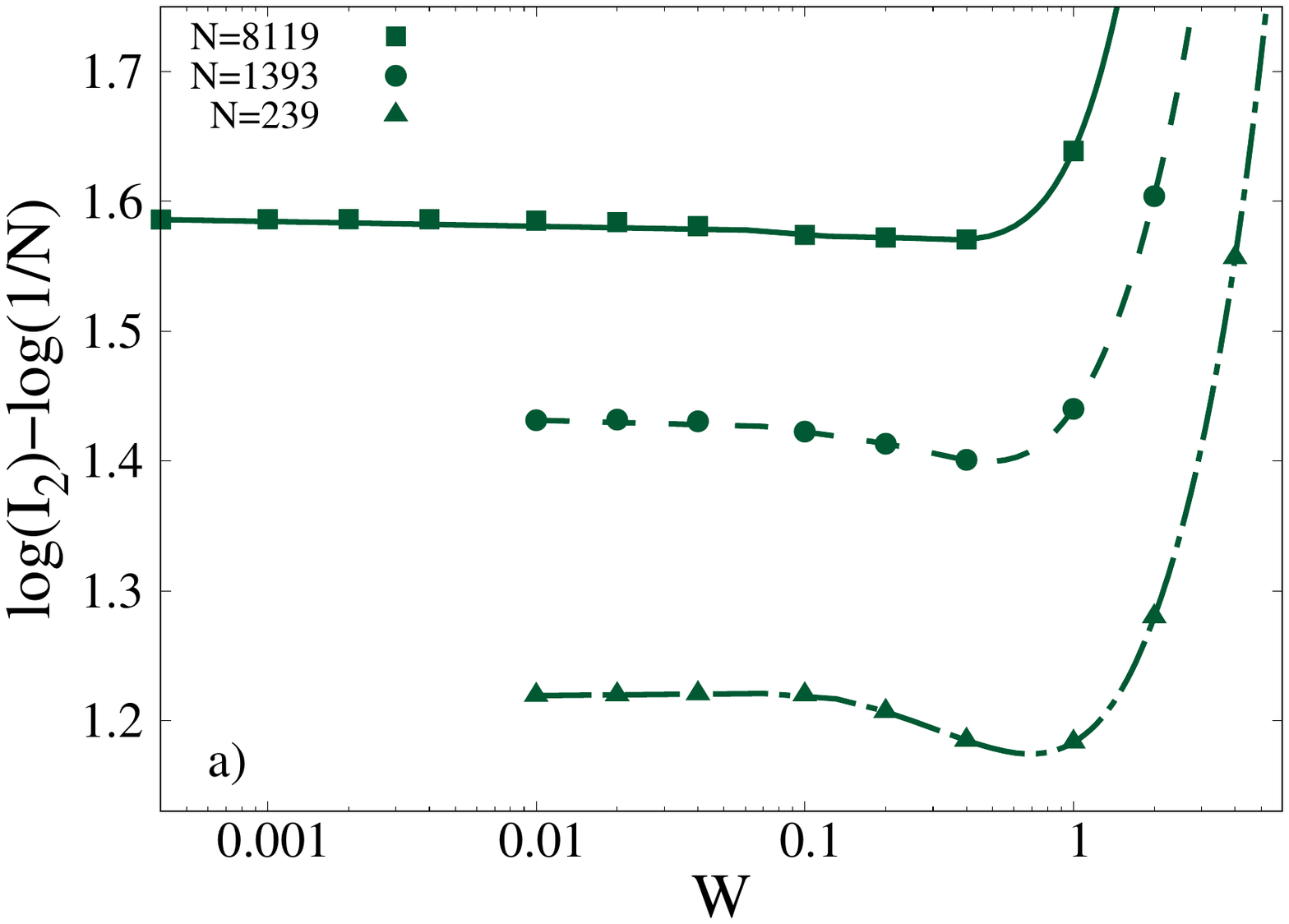} \hspace{0.1cm}
\includegraphics[width=0.431\textwidth]{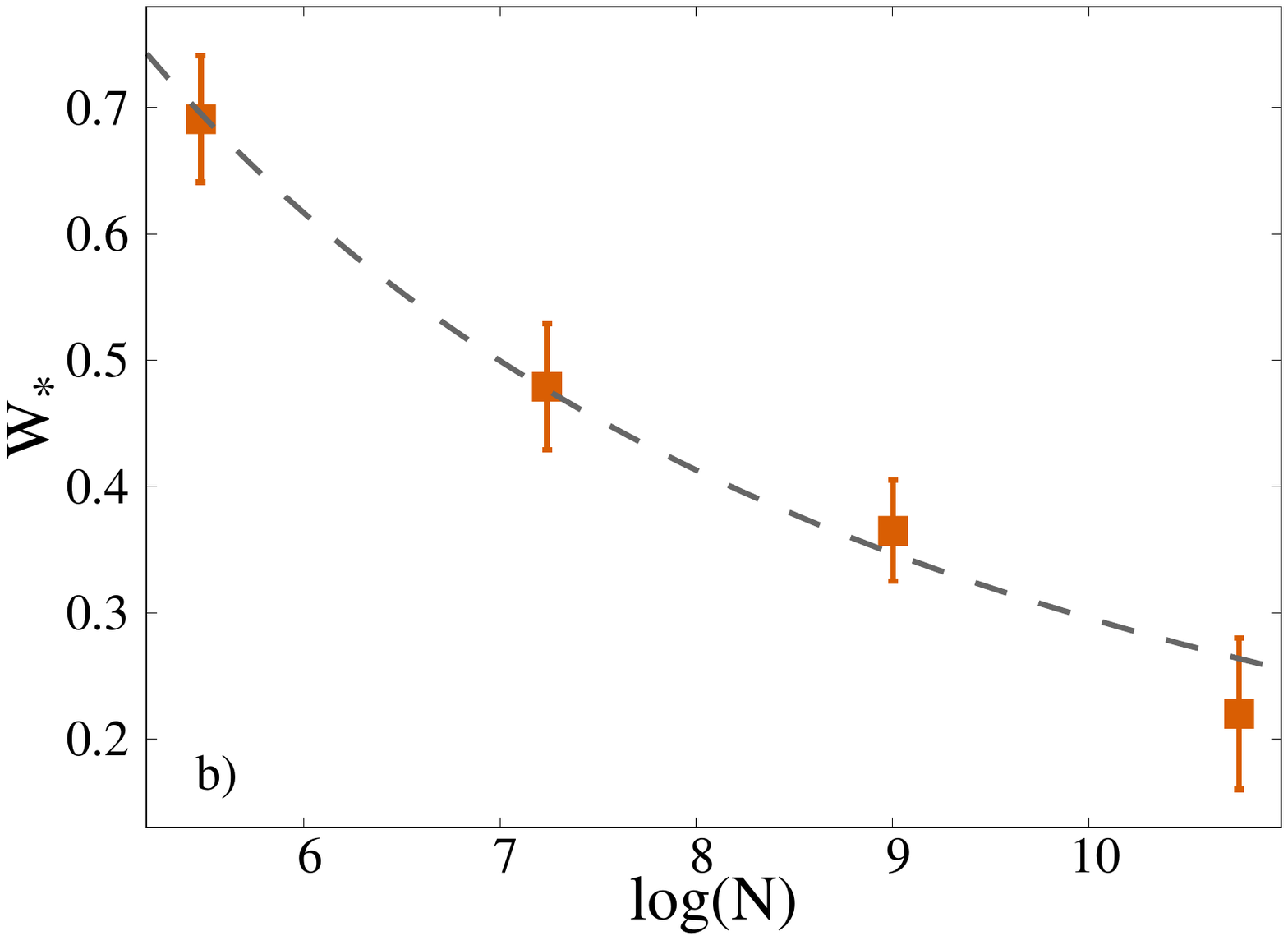} 
\caption{a) Size dependence of the $I_2$ plotted {\it vs} $W$ for $E=0.447$. b) Characteristic crossover scale $W_\star$ (estimated from the position of the minima of the data for $E=0.447$) as a function of the system size. The dashed line is a fit of the data to the form written in Eq.~\eqref{eq:wstar}. A similar dependence on $N$ is found for other values of $E$ in the bulk of the spectrum (i.e., away from the pseudogap and the edges of the band). 
\label{fig:ipr2}}
\end{figure}

The position of the minimum of $I_2$ depends on the energy, and varies in an irregular way, signaling the fact that the responsiveness of different eigenstates to the random perturbation fluctuates strongly. For $N=8119$, we find that $W_\star$ can vary between $0.07$ and $0.28$ when varying $E$ across the energy band. 
Fig.~\ref{fig:ipr2}a) shows a plot of $I_2$ {\it vs} $W$ for $E=0.447$  for three values of $N$. This plot shows that 
the minimum of $I_2$ is shifted to larger values of the disorder for smaller sizes. This behavior can be understood as follows: The disorder scale $W_\star$ corresponds to the value of $W$ such that the 2D localization length $\xi_{\rm loc}$ becomes of the order of the system size $L = N^{1/2}$. Since in 2D $\xi_{\rm loc}$ diverges exponentially for $W \to 0$, 
the larger is $N$ the smaller is $W_\star$. 
More precisely, one expects that :
    \begin{equation} \label{eq:wstar}
    W_\star \propto \left( \log N - c(E) \right)^{-2} \, .
    \end{equation}
where $c(E)$ is an energy-dependent parameter which  depends on the 
eigenvectors' sensitivity to the random perturbation, 
and fluctuates strongly from one eigenvector to another. This scaling of $W_\star$ is verified in Fig.~\ref{fig:ipr2}b), where we plot the evolution of $W_\star$ as a function of the log of the system size for $E=0.447$ (a similar trend with $N$ is found for other values of $E$ in the bulk of the spectrum).  The dashed line is a fit of the data of the form of Eq.~\eqref{eq:wstar}, which indeed reproduces well the results. Since D $=2$ is the lower critical dimension for Anderson localization, in our 2D model, the characteristic crossover scale $W_\star$ decreases exceedingly slowly with the system size, and is still quite large even for very large approximants.

{\it{States near the band center:}} It is interesting to examine in detail the behavior of confined states at $E=0$ of the pure model when disorder is added. Fig.~\ref{fig:ipr3}b) shows a close-up of the data of Fig.~\ref{fig:ipr1}a) near $E=0$. One can see that the non-monotonicity of the IPR is very pronounced for $E\approx 0$.  Consider states of energy exactly at $E=0$:  for the pure system their IPR has a large value ($I_2 \approx 0.1$), which, as the disorder is increased from $W=0$ to $W=2$, rapidly decreases by two orders of magnitude down to $I_2 \approx 10^{-3}$. At $W=2$, the confined states have an IPR of the same order of magnitude as band states in this energy range. For stronger disorder, beyond $W=2$, the IPR shows a rapid increase. The reason for this strong non-monotonicity of $I_2$ is quite evident. The degenerate localized states at $E=0$ in the pure system gives rise to high IPR values at $W=0$. When the diagonal random perturbation is turned on, the degeneracy is lifted, the energies of these eigenstates are scrambled around in a window of width $W$. At small $W$ the on-site random energies have the effect of delocalizing the eigenstates,  resulting in a rapid decrease of their IPR by more than two orders of magnitude.  Upon increasing the disorder the IPR continues to decrease until $W\approx 2$, and then start to increase rapidly due to standard Anderson localization at larger $W$. Also interesting to note for weak disorder are the maxima, or bumps, in the IPR curves in Fig.~\ref{fig:ipr1}a), which move to higher energies as disorder is increased.  The bumps arise due to realizations of the disorder in which the random energies $\epsilon_i$ have a nonzero positive or negative average value. For these realizations of the disorder the perturbation shifts the energy, but the overlaps with other states are small. This leads to the peak of the IPR for $E\sim W$. This phenomenon is not particular to quasicrystals and can be seen (as we have checked \cite{ajunpub}) in other systems with localized states such as the dice lattice \cite{sutherland}.

{\it{IPRs near the pseudogap and the band edge:}} Fig.~\ref{fig:ipr3}c) shows a close-up of the IPR data in the region close to the pseudogap at $E \approx 1.95$. Here non-monotonic behavior seems to be absent, within our numerical accuracy. Yet, for these states too, there is a characteristic crossover scale $W_\star$ such that for $W \lesssim W_\star $ disorder has very little effect on the IPR, while for $W > W_\star$ the IPR grows rapidly. For the system size shown in Fig.~\ref{fig:ipr3}c), $W_\star \sim 0.1$. Fig.~\ref{fig:ipr3}b) shows that a large peak of IPR develops for moderate values of $/W$ near the pseudogap. This behavior -- the absence of non-monotonicity and the appearence of the peak -- could arise due to an effect commonly seen at the band edge: Studies of Anderson localization in many models show that localization is strongly enhanced close to a band edge, which is what the pseudogap is in a certain sense.\footnote{Indeed a true gap can be opened at this location by  adding a small potential energy term to the hopping Hamiltonian.}  A similar monotonic increase of the IPR as a function of $W$, accompained by the development of a strong peak,  is also observed at the spectrum edge, $E \approx 4.2$ (see Fig.~\ref{fig:ipr1}a).
This type of behavior of the states near the pseudogap and the band edge has been observed also in the disordered 1D Fibonacci chain  \cite{disorderFC1,disorderFC2}. In fact, whereas non-monotonicity appears to hold as a general rule in 2D, it is observed only for a small subset of states in the 1D case. In 1D, most states localize monotonically like band edge states -- and indeed for the 1D quasicrystal, gaps are present throughout the spectrum.

\section{The anomalous dimensions of the wave-function amplitudes and the singularity spectrum} \label{sec:multi}

The non-monotonic crossover of the generalized inverse participation ratios of the wave-functions can be characterized in a more quantitative way by computing the flowing fractal dimensions $D^\psi_q$ associated to the scaling of the moments of the wave-functions' amplitudes. In order to do this we follow the approach of Ref.~\cite{slevin} which we detail below: The first step consists in covering the lattice with $N_l$ boxes of linear size $l$. The probability to find the particle in the $k$-th box ${\cal B}_k$ is simply given by:
\[
\mu_k(l) = \sum_{r \in {\cal B}_k} |\psi(r)|^2 \, , \qquad k = 1, \ldots, N_l \, .
\]
The $\mu_k(l)$ constitutes a normalized measure for which we can define the $q$-th moment as
\[
\upsilon_q(l) = \sum_{k=1}^{N_l} \mu_k^q(l) \, .
\]
The moments $\upsilon_q$ can be considered as the generalized inverse-participation ratios $I_q$, defined in Eq.~\eqref{eq:Pq}, for the integrated measure $\mu_k(l)$. 
The general assumption underlying multifractality is that within a certain range of values for the ratio $\zeta = n / N$, where $n$ is the number of sites within the boxes of size $l$, the moments $I_q$ show a power law behavior indicating the absence of length scales in the system,
\begin{equation} \label{eq:tau}
\upsilon_q(\zeta) \propto \zeta^{\tau_q} \, ,
\end{equation}
with the mass exponents $\tau_q$ defined in Eq.~\eqref{eq:tau}.

\begin{figure}[h]
\includegraphics[width=0.6\textwidth]{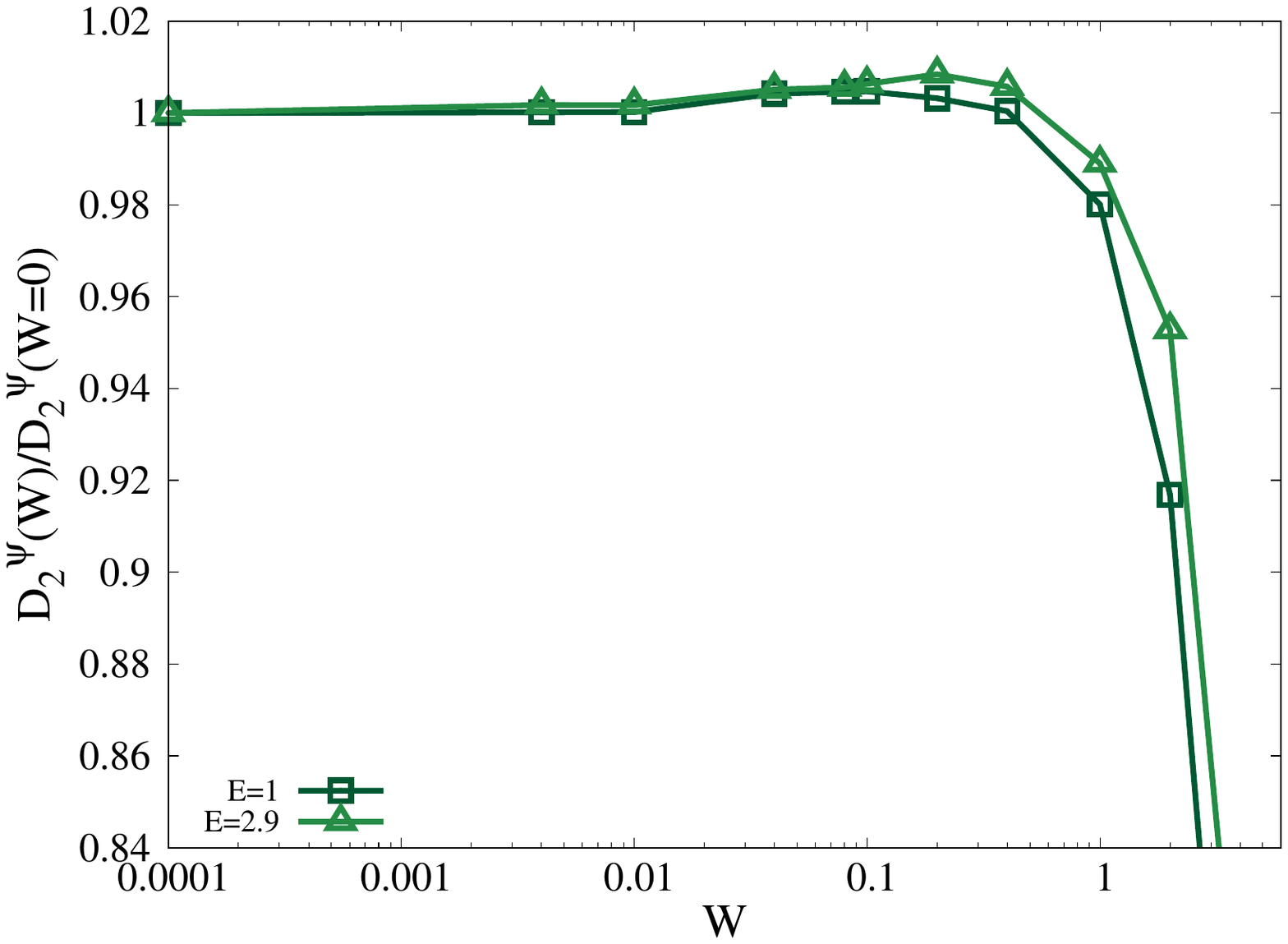} 
\caption{Relative change of the fractal dimension, $D^\psi_2(W)/D^\psi_2(0)$ plotted against $W$ for two different energies, showing their non-monotonicity. (Numerical estimations based on Eq.~\eqref{eq:tauq} for $N=47321$.)}
\label{fig:D2}
\end{figure}

The numerical analysis is essentially based on an averaged form of the scaling law Eq.~\eqref{eq:tau} in the limit $\zeta \to 0$. The scaling exponents are then computed as:
\begin{equation} \label{eq:tauq}
\tau_q = \lim_{\zeta \to 0} \frac{\left \langle \log \upsilon_q(\zeta) \right \rangle}{\log \zeta} \equiv (q-1) D_q^\psi \, ,
\end{equation}
and can be obtained from the extrapolation at small $\zeta$ of the slope of the linear fit of $\langle \log \upsilon_q(\zeta) \rangle$ versus $\log \zeta$.\footnote{We have checked numerically that using the logarithm of the average moments $\log \langle \upsilon_q(\zeta) \rangle$ instead of the typical ones $\langle \log \upsilon_q(\zeta) \rangle$ yields essentially the same results but with larger statistical errors.} Here $\langle \cdots \rangle$ denotes the arithmetic average over many independent realizations of disorder. The numerical results for $D^\psi_2 = \tau_2$ obtained for the biggest tiling $N=47321$ using this procedure are shown in Fig.~\ref{fig:D2}. The figure shows the ratio $D^\psi_2(W)/D^\psi_2(0)$ plotted against $W$, for two different values of the energy  (similar results are found for other values of $q$ and $N$). This plot indicates that $D^\psi_2$ has a non-monotonic behavior as a function of $W$, and reaches a maximum around the characteristic disorder scale $W_\star$ which corresponds to the value of $W$ such that the 2D localization length $\xi_{\rm loc}$ becomes of the order of the system size $L = N^{1/2}$, Eq.~\eqref{eq:wstar}, and is consistent with the one found in the right panel of Fig.~\ref{fig:ipr3}. 
As said earlier in Sec.~\ref{sec:ipr}, the position of $W_\star$ at fixed $N$ depends slightly on $E$ and varies in an irregular way across the energy band.
 In the $W \to 0$ limit $D^\psi_2$ is quite close to $1$ and depends only weakly on $E$ -- thereby confirming that in the pure limit the eigenstates are critical but fairly extended. 
Above the crossover scale $W_\star$, $D^\psi_2$ decreases very sharply and approaches $0$ very fast, as the system enters in a strongly localized regime due to standard Anderson localization.
Although the non-monotonic behavior is more or less pronounced depending on the value of $E$ (as discussed above the non-monotonicity is very pronounced for $E \approx 0$), it seems that most of the 
states exhibit a maximum of $D^\psi_2$. However, as seen for the IPR, the non-monotonic behavior of $D_2^\psi$ disappears in the region of the pseudogap, and sufficiently close to the band edges.

\begin{figure}
\includegraphics[width=0.431\textwidth]{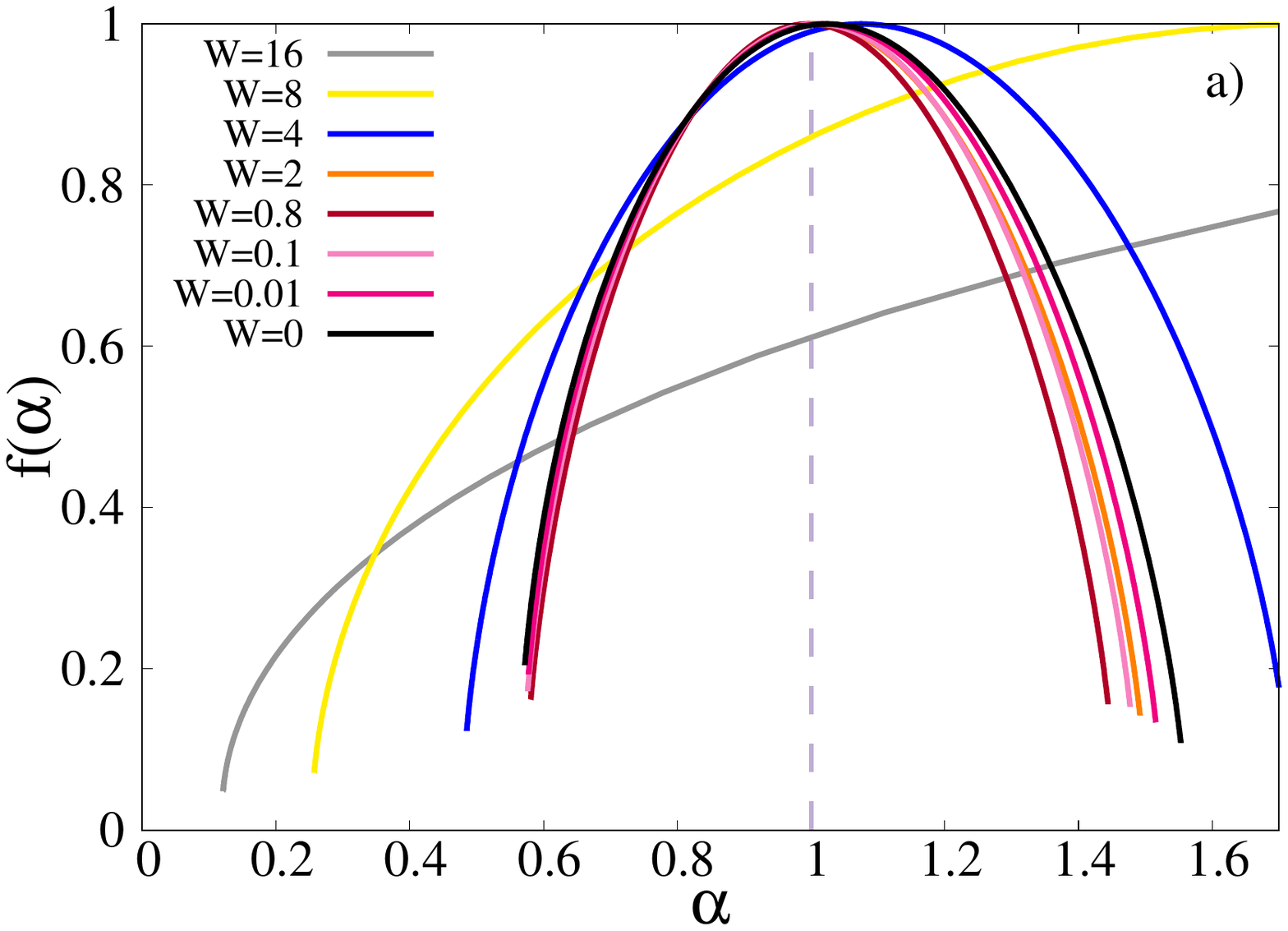} \hspace{0.1cm} \includegraphics[width=0.431\textwidth]{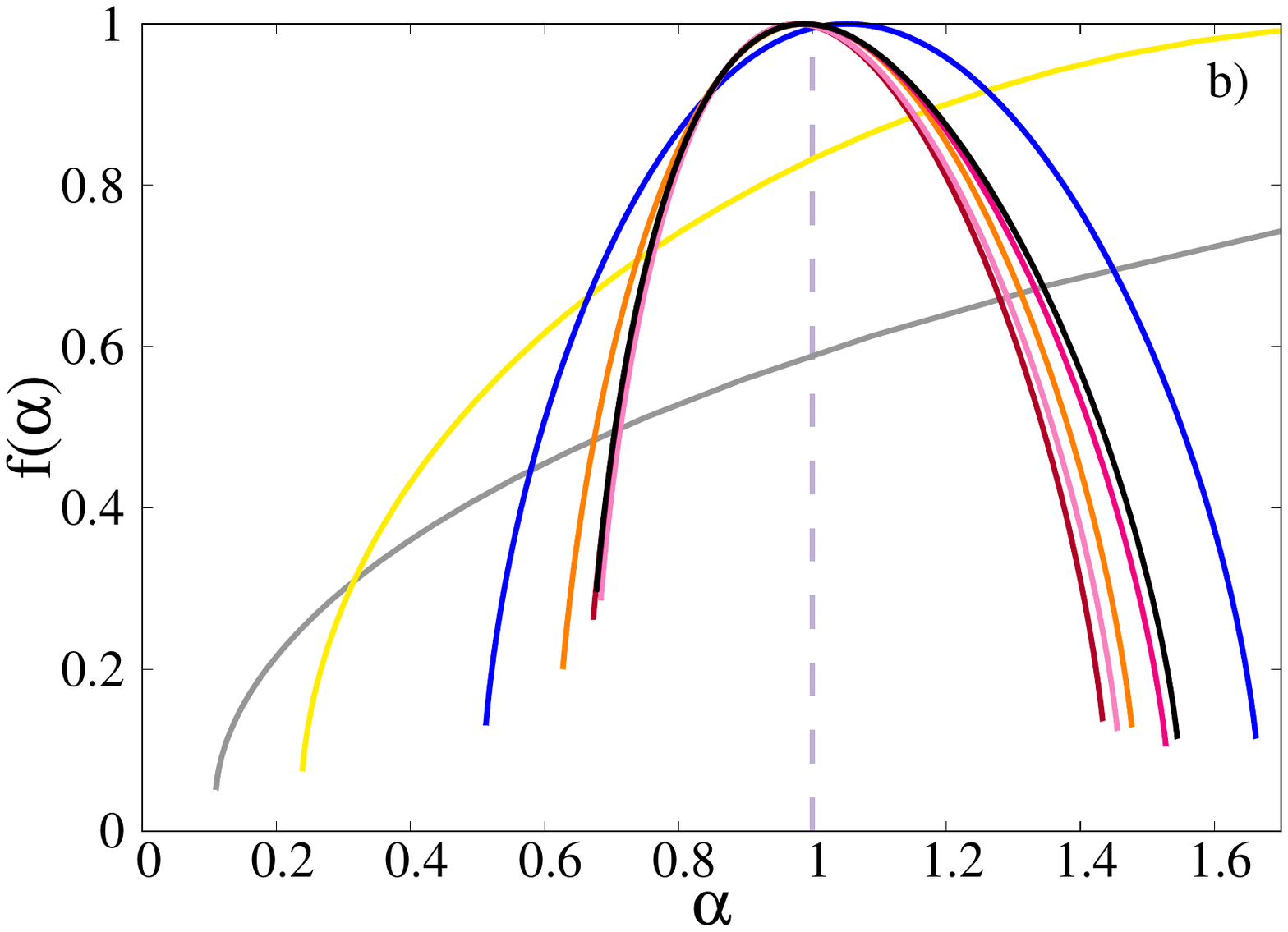}

\includegraphics[width=0.431\textwidth]{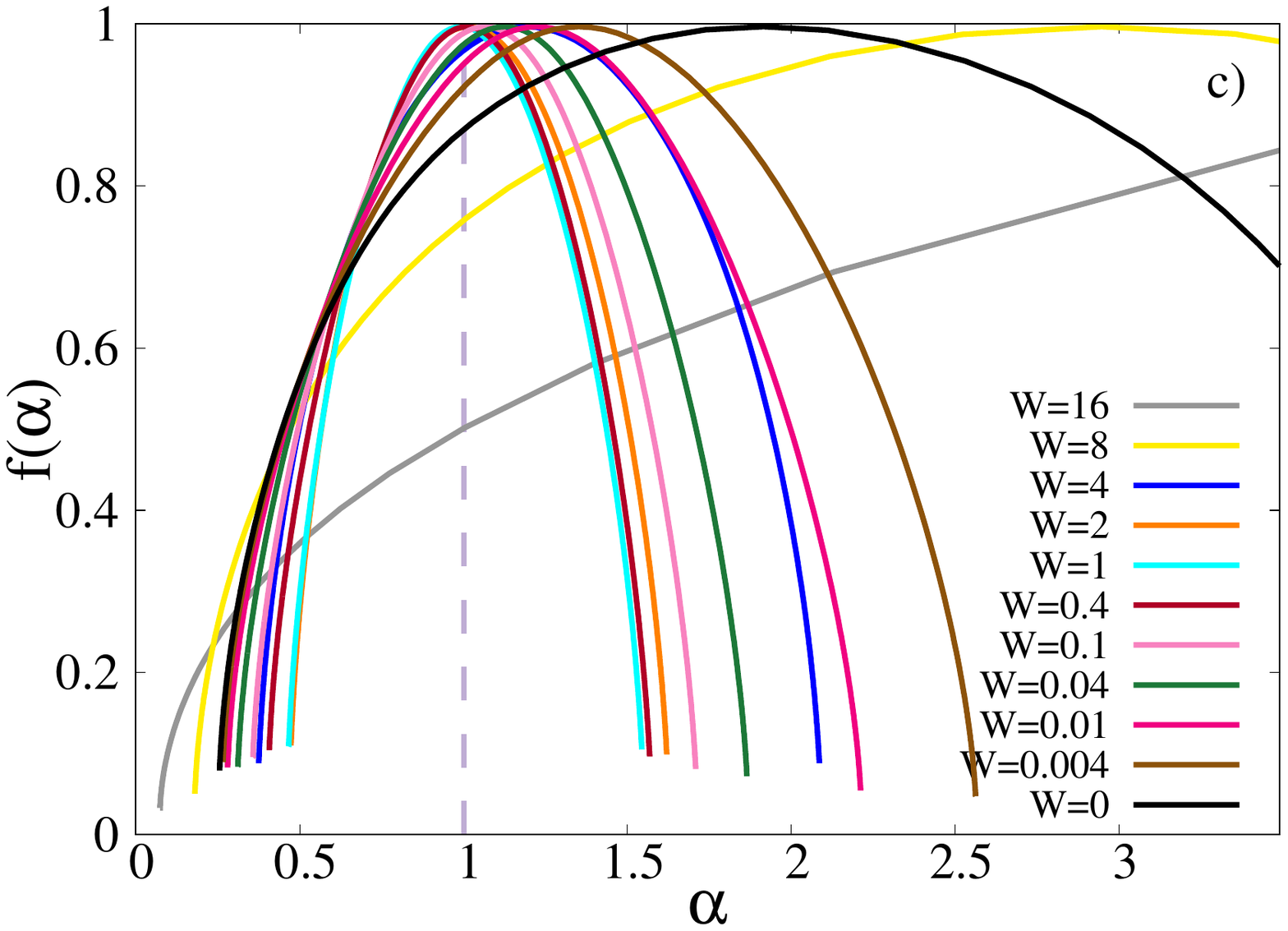} \hspace{0.1cm} \includegraphics[width=0.431\textwidth]{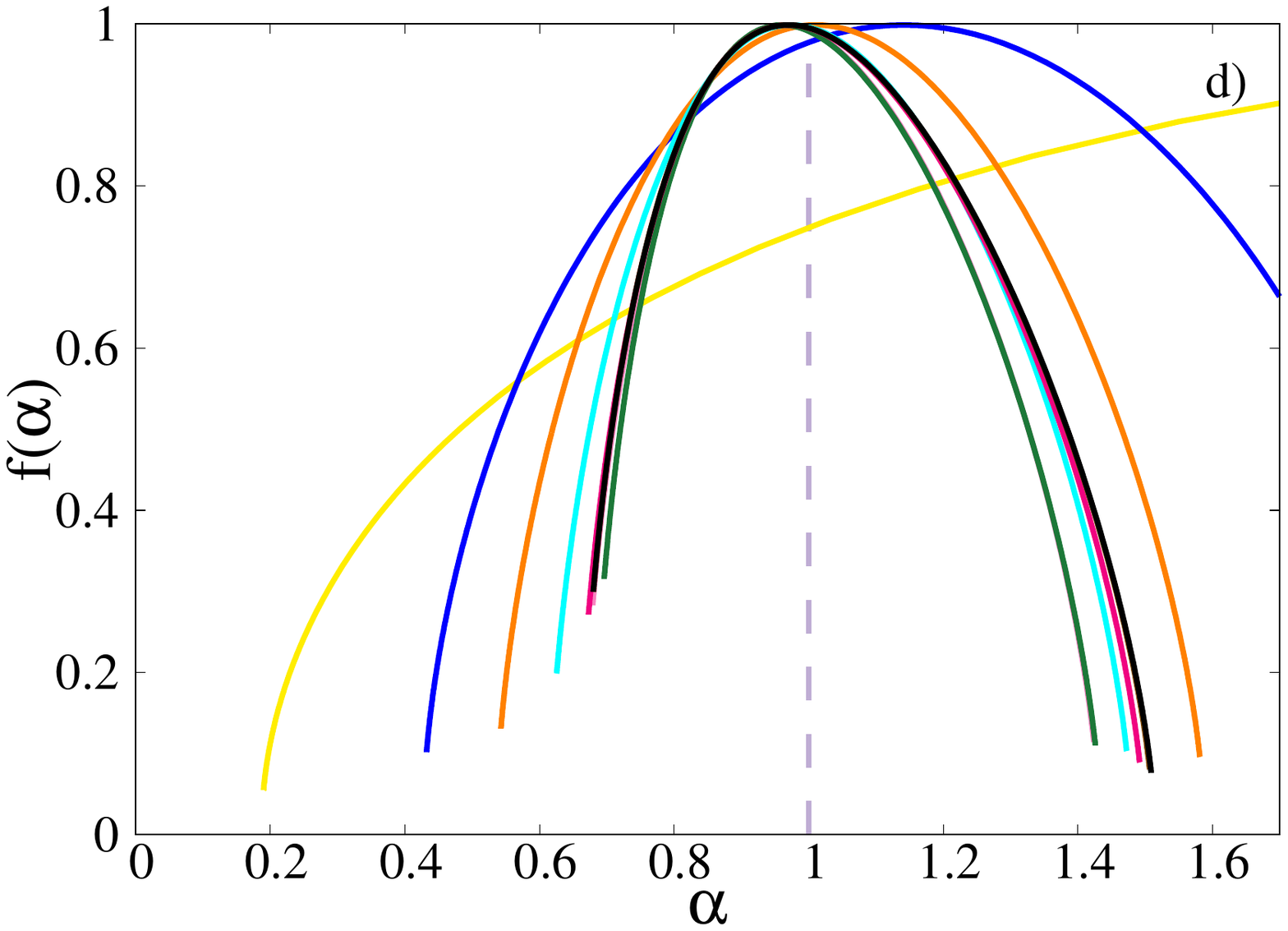}

\caption{Singularity spectrum $f(\alpha)$. Different colors correspond to different values of $W$, as indicated in the legends. The vertical dashed lines correspond to the singularity spectrum in the limit of fully delocalized eigenstates, $f(\alpha) = \delta(\alpha-1)$. a) Eigenstates considered are a) $E=0.74$, b) $E=2.25$, c) $E=0$ and d) $E=1.9$. System sizes are $N=8119$ (a and b) and $N=47321$ (c and d). The multifractal spectrum is the narrowest for $W \simeq 0.8$ in panels a) and b), for $W \simeq 1$ in panel c), and for $W \simeq 0.04$ in panel d).
\label{fig:multi}}
\end{figure}

From the exponents $\tau_q$ one can also compute the singularity spectrum $f(\alpha)$, defined as a Legendre transformation of the mass exponents $\tau_q$:
\[
\begin{aligned}
\alpha_q &= \frac{\de \tau_q}{\de q} \, , \qquad q = \frac{\de f}{\de \alpha} \, , \\
f_q & \equiv f(\alpha_q) = \alpha_q q - \tau_q \, .
\end{aligned}
\]
Here, $f(\alpha)$ denotes the fractal dimension of the set of points where the wave-function intensity is $|\psi (r)|^2 \propto N^{- \alpha}$, i.e. the number of such points scales as $N^{f(\alpha)}$. The singularity spectrum $f(\alpha)$ is a convex function of $\alpha$, and it has its maximum at $\alpha_0 \ge 1$, where $f(\alpha_0)=1$. One also has that $f(\alpha_1)=\alpha_1$ and $f^\prime (\alpha_1)=1$. For fully delocalized wave-functions the singularity spectrum becomes narrower and eventually converges to one point $f(1)=1$. On the other hand, for fully localized wave-functions the singularity spectrum broadens, and tends to converge to the points $f(0)=0$ and $f(\infty) = 1$. Numerically the singularity spectrum is obtained as~\cite{slevin}:
\[
\begin{aligned}
\alpha_q = \lim_{\zeta \to 0} \frac{1}{\log \zeta} \left \langle \sum_{k=1}^{N_\zeta} \delta_k (q,\zeta) \log \delta_k (1,\zeta) \right \rangle \, , \\
f_q = \lim_{\zeta \to 0} \frac{1}{\log \zeta} \left \langle \sum_{k=1}^{N_\zeta} \delta_k (q,\zeta) \log \delta_k (q,\zeta) \right \rangle \, , 
\end{aligned}
\]
where $\delta_k (q,\zeta) = \mu_k^q(\zeta)/\upsilon_q(\zeta)$ is the normalized $q$-th power of the integrated measure $\mu_k(\zeta)$.

The numerical results for $f(\alpha)$ are shown in Figs.~\ref{fig:multi}, where we plot the multifractal spectrum for many different values of the disorder, for the approximants with $N=8119$ (a and b) and $N=47321$ (c and d), and for four values of the energy across the spectrum as indicated in the caption. 
The non-monotonic behavior of the generalized inverse-participation ratios and of the fractal exponents discussed above is clearly reflected in the evolution of $f(\alpha)$ with $W$: At small enough disorder, $W < W_\star$, $f(\alpha)$ first slightly shrinks when the disorder is increased, indicating that the eigenstates tends to become more delocalized, and then starts to broadens again rapidly for $W > W_\star$, when the system enters in the strong disorder regime dominated by standard Anderson localization. The information contained in  Figs.~\ref{fig:multi} can be summarized as follows:

\begin{itemize}
\item {\it{Typical band states}}. Figs.~\ref{fig:multi}a) and b) correspond to two values of $E$ in the bulk of the spectrum, $E=0.74$ and $E=2.25$ (similar results are found for other values of $E$ far from the pseudogap or the spectral edges, and far from the middle of the band), and for $N=8119$. For this choice of the parameters $f(\alpha)$ is the narrowest for $W_\star \in [0.4 , 0.8]$. 
\item {\it{Confined states}}. Fig.~\ref{fig:multi}c)  shows the singularity spectrum of the eigenstates close to the middle of the band, $E=0$ (and for $N=47321$). As we have pointed out, for these states, the non-monotonic crossover is much stronger. In fact, since these eigenstates are genuinely localized in the pure limit, $f(\alpha)$ is quite broad for $W=0$. The effect of the addition of the random perturbation is thus much more pronounced, as it produces a strong shrinkage of the singularity spectrum which is the narrowest for $W_\star \sim 2$. For larger values of the disorder $f(\alpha)$ rapidly gets broader again as the system enters in the strong Anderson localized regime. This passage from ``confined'' state to band states has been observed in a different context, when a tiling is subjected to a perpendicular magnetic flux. Depending on the local geometry, there can be confined states, for special values of the magnetic flux, and these deconfine when the flux is modified, with the corresponding changes of their multifractal spectrum as computed in \cite{fluxmodel}.
\item {\it{Pseudogap states}}. Figs.~\ref{fig:multi}d)  shows the results for the eigenstates close to the pseudogap, $E=1.9$ (and for $N=47321$), for which the non-monotonic behavior is either absent of very weak -- see  Fig.~\ref{fig:ipr3}b). The plot shows that the left branch of $f(\alpha)$ (for $\alpha<1$) is essentially independent of $W$ at small disorder, and only the right branch of the curve exhibit a slight narrowing as $W$ is increased up to $W_\star$. For larger disorder once again we observe that the singularity spectrum approaches rapidly the one corresponding to the strongly localized regime.
\end{itemize}

\section{Eigenstates correlations} \label{sec:corr}

An important statistical characteristic of a disordered system is the spatial correlations of different (but relatively close in energy) eigenstates, defined as:
\[
C(r,\omega;E) \equiv \frac{1}{N^2 \rho \left (E- \frac{\omega}{2} \right) \rho \left (E+ \frac{\omega}{2} \right)} \left \langle \sum_{i,j} | \psi_i (r_1) \psi_j (r_2) |^2 \delta\left( E - \frac{\omega}{2} - E_i\right) \delta\left( E + \frac{\omega}{2} - E_j\right) \right \rangle \, .
\]
where $r=|r_1-r_2|$ is the spatial distance and $\omega = E_1-E_2$ is the energy separation between the states. The denominator counts the number of pairs of states entering in the sum in the numerator.
$\langle \cdots \rangle$ here denotes the average over several independent realizations of the disorder, and all pair of sites of the lattice at fixed distance (i.e. we assume translation and rotation invariance on average). This correlation function encodes a lot of information about the structure of the wave-functions and the spectral statistics. To simplify the analysis in the following we start by studying correlations of the same wave-function (i.e. $\omega=0$) at different points in space, and of wave-functions at different energies but at the same point in space (i.e. $r_1=r_2$). Once the behavior of these objects is established, we will also address the correlations of the amplitudes of different eigenstates taken at different points of the lattice.

\subsection{Same wave-function, different positions}

\begin{figure}
\includegraphics[width=0.331\textwidth]{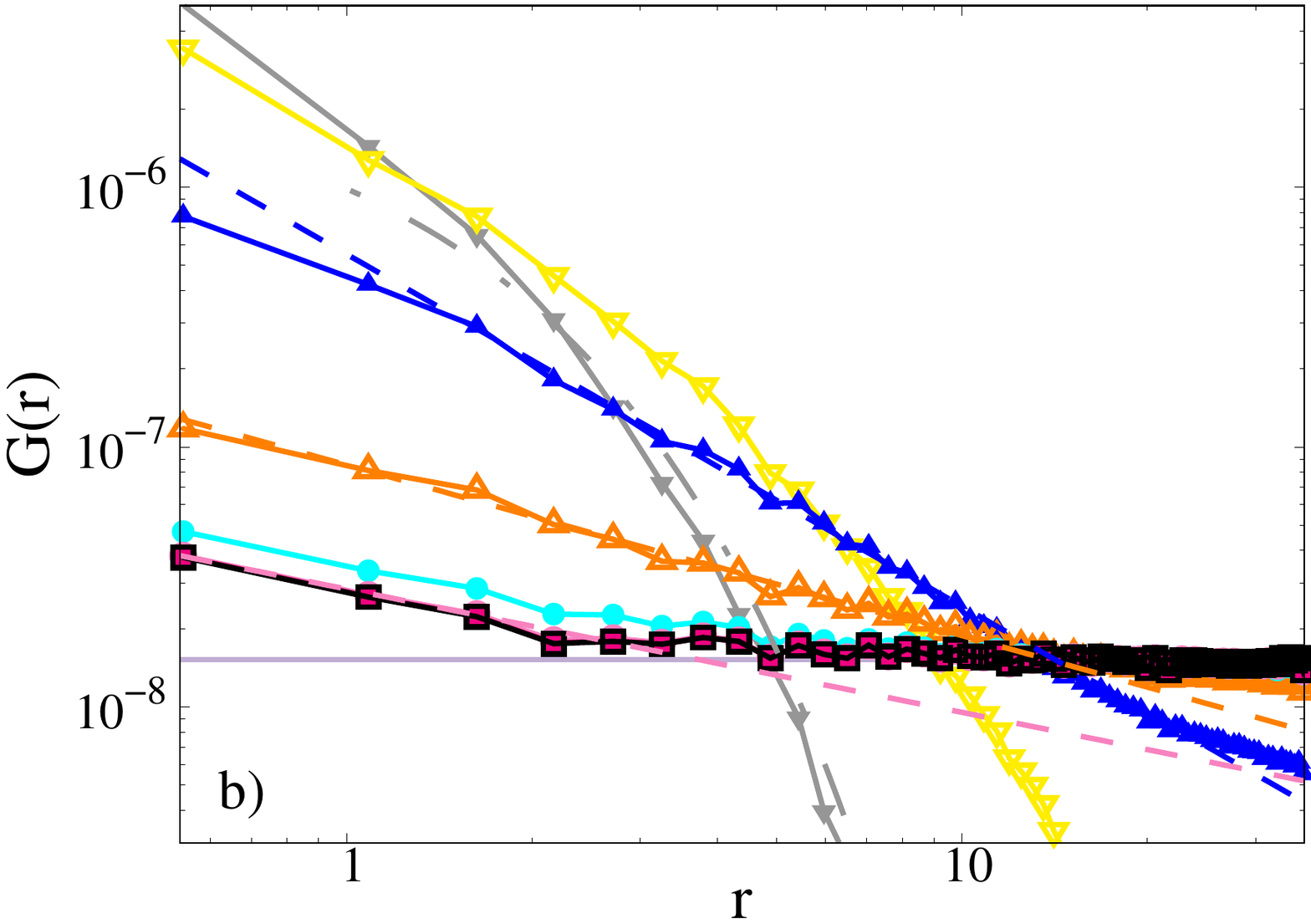} \hspace{-0.1cm} \includegraphics[width=0.331\textwidth]{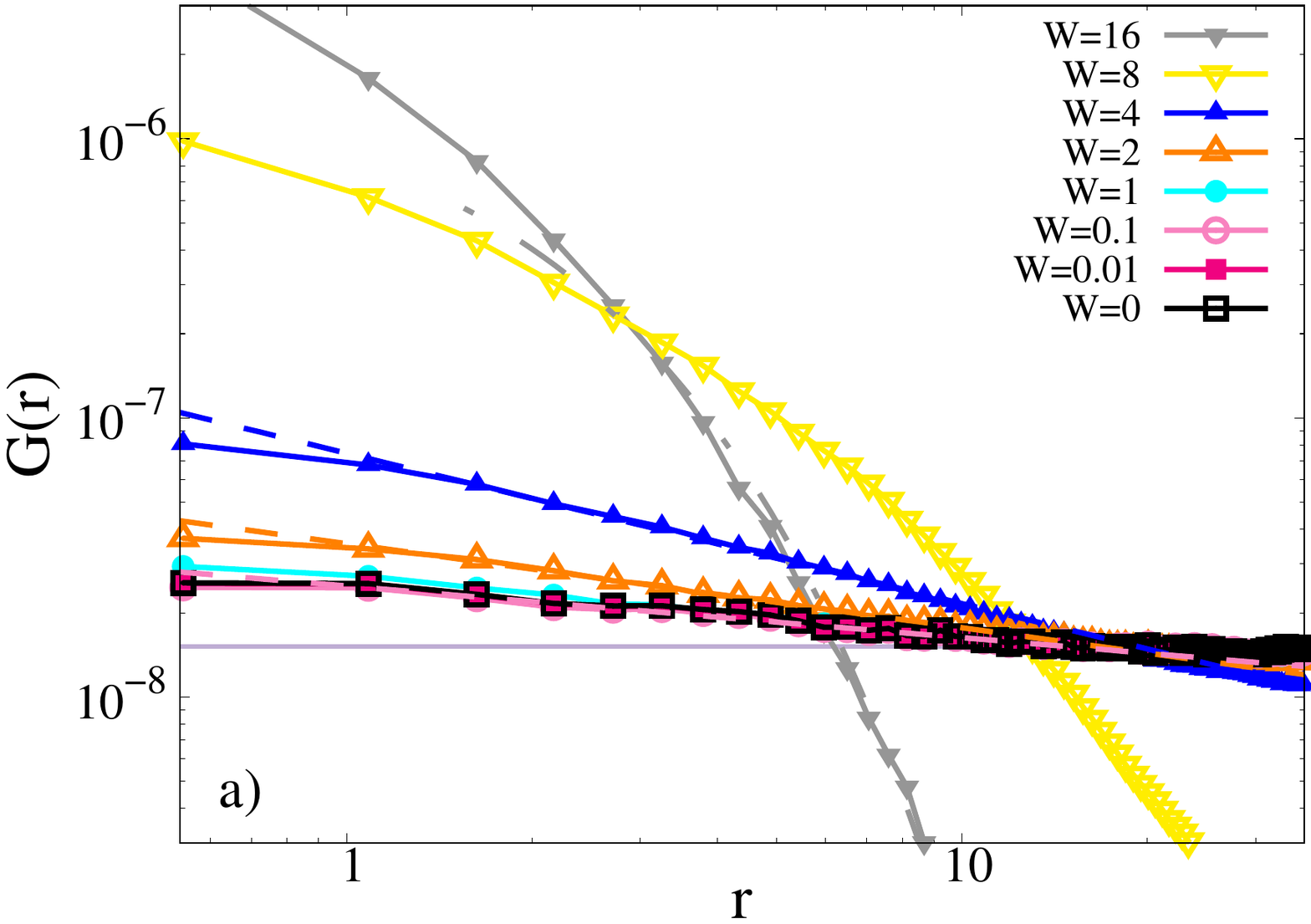} \hspace{-0.1cm}
\includegraphics[width=0.331\textwidth]{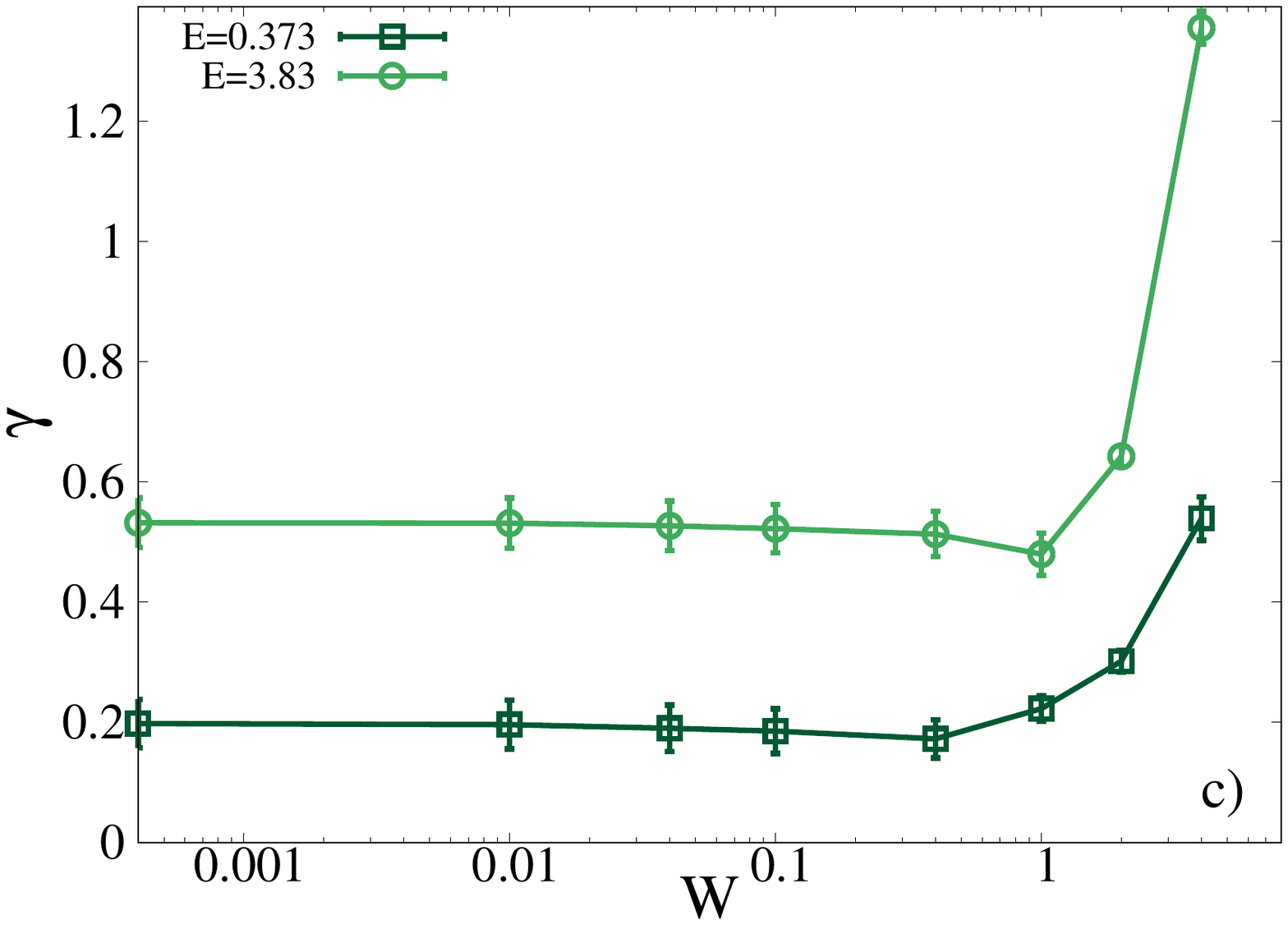} 

\vspace{0.2cm}

\includegraphics[width=0.331\textwidth]{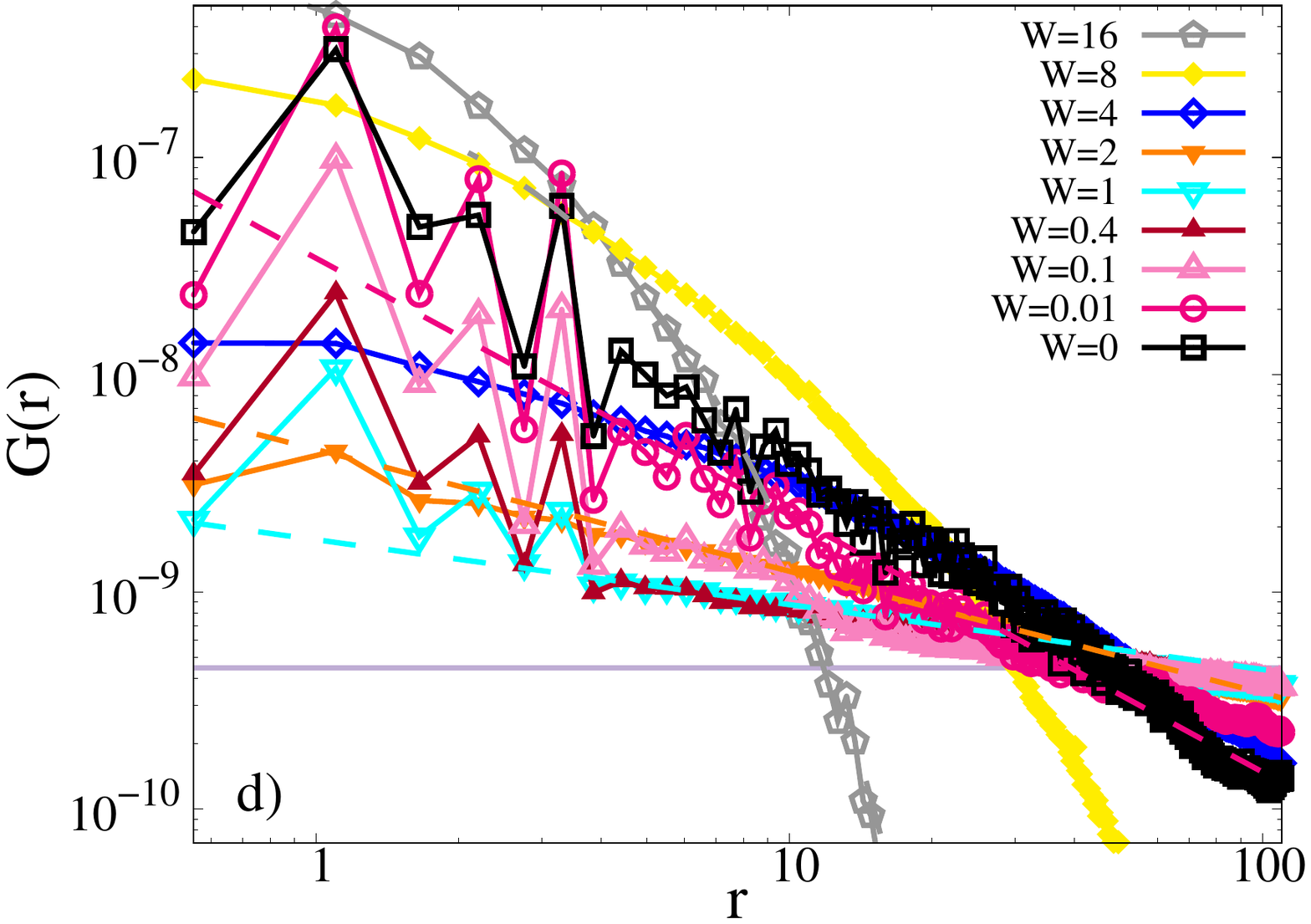} \hspace{-0.1cm} \includegraphics[width=0.331\textwidth]{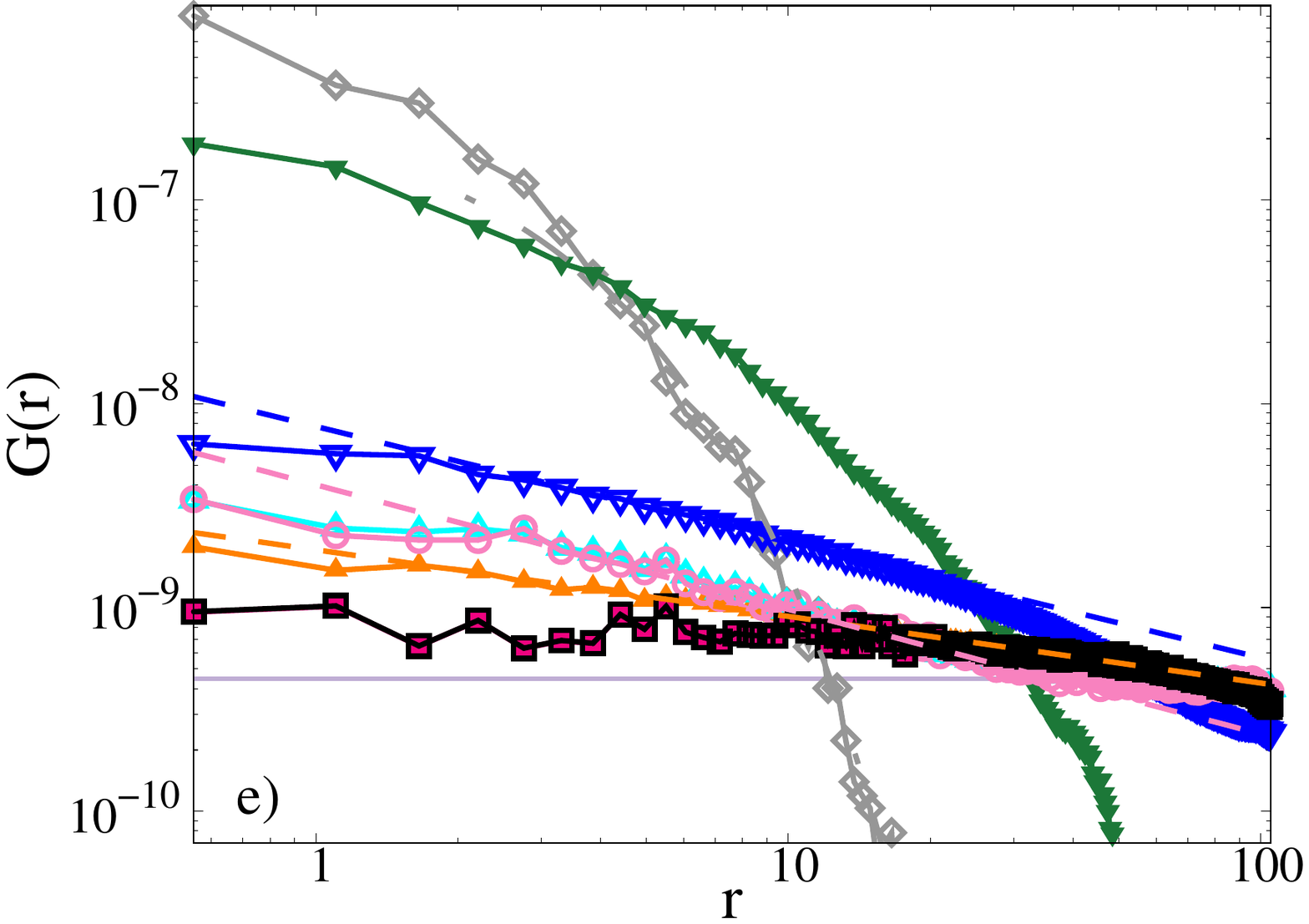} \hspace{-0.1cm}
\includegraphics[width=0.331\textwidth]{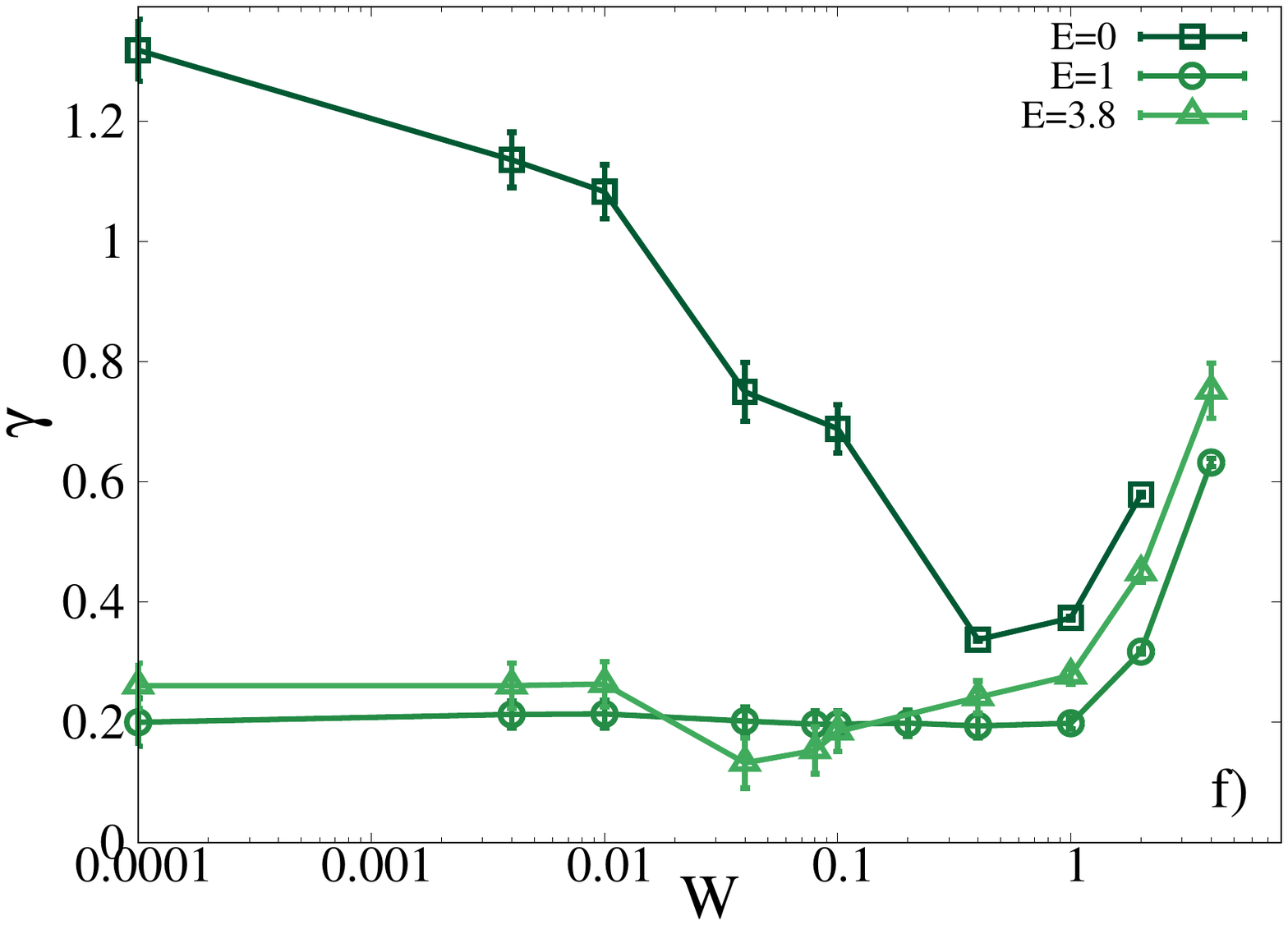}
\caption{Log-log plot of the correlation function $G(r;E)$ for several values of the disorder and for two system sizes. Panels a and b  show results for $N=8119$ for the energies  $E=0.373$ (a) and $E=3.8$ (b). Panels d and e shows result for
$N=47321$ for $E=0$ (d), and $E=1.9$ (e). The dashed straight lines are power law fits of the data of the form $G(r;E) \sim r^{- \gamma}$, with a disorder-dependent effective exponent $\gamma$. The dashed-dotted lines are exponential fits of the form $G(r;E) \sim e^{-r / \xi_{\rm loc}}$. The horizontal full line corresponds to $1/N^2$. In panels c and f the exponent $\gamma$ extracted from the fits of the data of panels a, b, d, e 
is plotted as a function of $W$, showing a non-monotonic behavior on the characteristic crossover scale $W_\star$ for $N=8119$ (c) and $N=47321$ (f).
\label{fig:corr}}
\end{figure}

We consider first the auto-correlation of the amplitudes of a given wave-function ($\omega = 0$) measured at different positions:
\[
G(r;E) \equiv  \frac{1}{N \rho (E) } \left \langle \sum_{i} | \psi_i (r_1) \psi_i (r_1 + r) |^2 \delta\left( E - E_i\right) \right \rangle  = \lim_{\omega \to 0} C(r,\omega;E)\, .
\]

Let us recall the behavior of $G(r;E)$ for some limiting cases. For fully delocalized eigenstates described by the GOE ensembles, $G(r;E) = 1/N^2$ identically, since the amplitudes $\psi_i$ correspond to random vectors on the unit sphere. For fully Anderson localized eigenstates, $G(r;E)$ decays exponentially over the localization length as $G(r;E) \sim e^{-r / \xi_{\rm loc}}$. Finally, for multifractal/critical eigenstates one expects that $G(r;E)$ decays algebraically, $G(r;E) \sim r^{- \gamma}$. Note that for $r=0$ the function simply boils down to the IPR, that is, $G (0;E) = I_2(E)/N$.

The numerical data for $G(r;E)$ are plotted in Fig.~\ref{fig:corr}. In panels a and b  we show the results for $N=8119$ and for two values of the energy in the bulk of the spectrum. In panels d and e we plot $G(r;E)$ for the largest available size $N=47321$, and for eigenstates in the middle of the band, $E=0$, and close to the pseudogap, $E=1.95$, respectively. We find essentially the same results for the approximants of size $N=8119$ and $N=47321$, although the latter are more noisy due to poorer statistics for this larger system. One sees that for all cases, at small enough disorder, $G(r;E)$ decays as a power law (dashed lines), with a disorder-dependent effective exponent $\gamma$.
For large disorder, instead, $G(r;E)$ decays exponentially (dashed-dotted lines). Fig.~\ref{fig:corr} shows, in turn, that the qualitative differences noted in the behavior of the IPR and of the multifractal spectrum in the previous section are also to be seen in these results for $G(r;E)$. We see again three types of behavior as follows:\\
\begin{itemize}
\item For eigenstates in the bulk of the spectrum (panels a and b), the effective exponent $\gamma$ has a non-monotonic behavior as a function of $W$ on the characteristic crossover scale $W_\star$: It first slightly decreases as $W$ is increased (corresponding to wave-functions that become more extended as the disorder is increased), and then it increases with $W$ for $W > W_\star$, until the power law decay transforms into an exponential decay. Such non-monotonic behavior is highlighted in panels c and f of the figure, where we plot the values of $\gamma$ extracted from the power law fits of $G(r;E)$ as a function of the disorder. 
\item For eigenstates close to the middle of the band the non-monotonic crossover is strongly enhanced  ($E=0$, panel d). For these states the exponent $\gamma$ decreases in a very pronounced way as $W$ is increased from $0$ to $1$ -- see Fig.~\ref{fig:corr}f). 
\item Non-monotonicity is not observable (upto numerical accuracy) for eigenstates close to the pseudogap ($E=1.95$, panel e), for which the exponent $\gamma$ seems to be independent of $W$ at small disorder but increases sharply after $W_*$.
\end{itemize}

\subsection{Same position, different energies} \label{sec:K2}
A valuable probe of the level statistics and of the statistics of wave-functions' amplitudes is provided by the spectral correlation function $K_2 (\omega;E)$ between eigenstates at different energy, but on the same site, which allows one to distinguish between ergodic, localized, and multifractal states~\cite{altshulerK2,mirlin,chalker,kravK2,kravtsov}:
\begin{equation} \label{eq:K2}
	\begin{aligned}
K_2 (\omega;E) & \equiv  \frac{\left \langle N \sum_{i,j} \sum_r | \psi_i (r) \psi_j (r) |^2 \delta\left( E - \frac{\omega}{2} - E_i\right) \delta\left( E + \frac{\omega}{2} - E_j\right) \right \rangle }{N^2 \rho \left (E- \frac{\omega}{2} \right) \rho \left (E+ \frac{\omega}{2} \right)} = N^2 C(E;0,\omega) \, .
\end{aligned}
\end{equation}
Let us begin by recalling the expected behavior of $K_2(\omega;E)$ for some well-studied cases. For random GOE matrices $K_2(\omega;E) = 1$ identically, independently of $\omega$ on the entire spectral bandwidth. On the other hand, in a standard metallic phase (e.g., in the extended phase of the Anderson tight-binding model in D~$\ge 3$) $K_2$ has a plateau at small $\omega$, for $\omega$ smaller than a characteristic energy scale  $\omega_{\rm AS}$, followed by a fast-decay which is described by a power law, with a system-dependent exponent~\cite{chalker}. The height of the plateau is larger than one, which implies an enhancement of correlations compared to the case of independently fluctuating Gaussian wave-functions. The energy scale $\omega_{\rm AS}$ is the so-called ``Altshuler-Shklovskii'' energy, defined as the Thouless energy at the scale of the lattice spacing ($\omega_{\rm AS} = D/a^2 = N E_{\rm Th}$, $D$ being the diffusion coefficient), which plays a role of the ultraviolet cut-off for the diffusion theory, separating the plateau region (corresponding to long times) and the the power law decay (for shorter times). The width of the plateau at $\omega \lesssim \omega_{\rm AS}$ remains finite in the thermodynamic limit and moves to larger energies as one enters further into the metallic phase, and corresponds to the width of the energy band over which one expects to find GOE-like correlations ~\cite{altshulerK2}. For multifractal states the plateau is usually present only in a narrow energy interval, and $\omega_{\rm AS}$ shrinks to zero in the thermodynamic limit, still staying much larger than the mean level spacing. Beyond $\omega_{\rm AS}$, eigenfunctions have less and less overlap with each other and the statistics is no longer Wigner-Dyson, and thus $K_2(\omega;E)$ decays to zero~\cite{kravtsov}. Note that in the limit of vanishing energy difference, the function simply boils down to the IPR, that is, $\lim_{\omega \to 0} K_2 (\omega;E) = N I_2(E)$.

\begin{figure}
\includegraphics[width=0.431\textwidth]{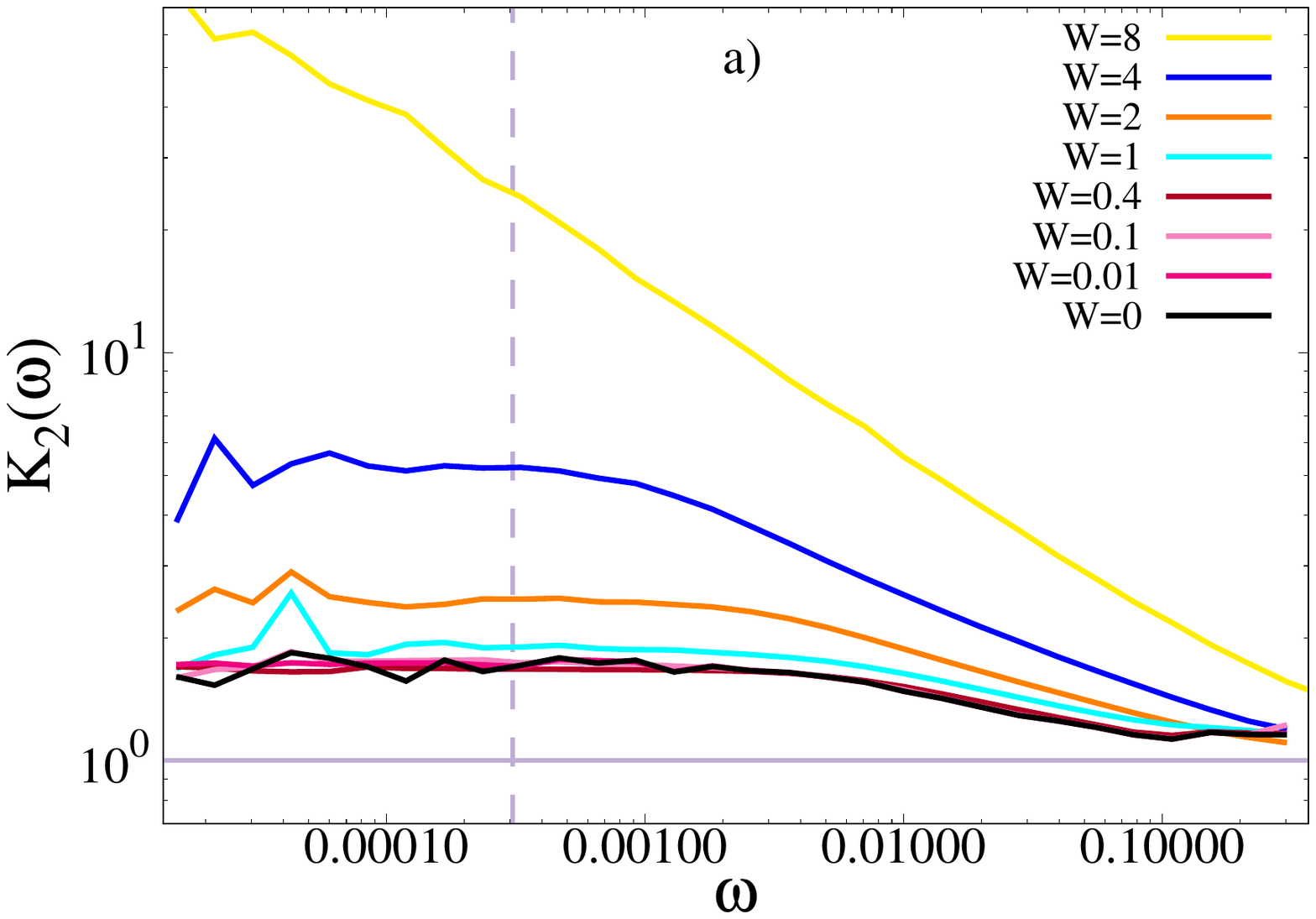}  \hspace{0.1cm} \includegraphics[width=0.431\textwidth]{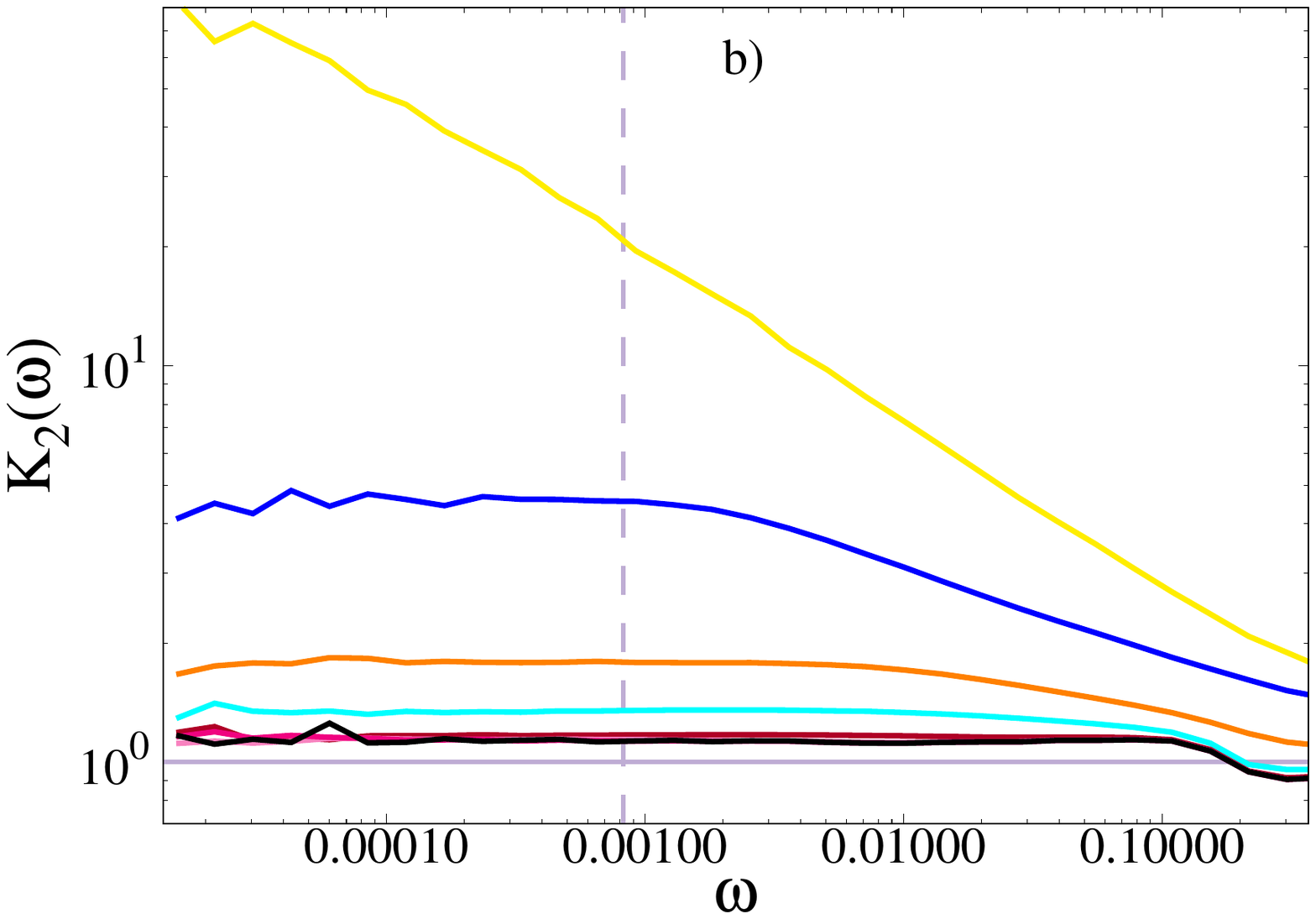} 
\caption{Overlap correlation function $K_2(\omega)$ for several values of the disorder, for $N=8119$, and for two values of the energy, $E=0.373$ (a) and $E=3.8$ (b). The horizontal line represents the GOE limit, $K_2(\omega;E)=1$, while the vertical dashed lines spot the value of the energy separation corresponding to the mean level spacing at $W=0$.
\label{fig:K2}}
\end{figure}

Fig.~\ref{fig:K2}  shows numerical results for the overlap correlation functions obtained for the approximant of size $N=8119$, for several values of the disorder strength $W$. The two plots correspond to two values of the energy chosen within the bulk of the spectrum $E=0.373$ (a) and close to the band edge, $E=3.8$ (b).  The crossover to localization clearly manifests itself in the behavior of $K_2 (\omega;E)$. In the pure limit, $W=0$, the plateau region shows small oscillations, which reflect the correlations between eigenstates belonging to different symmetry sectors~\cite{grimmroemer}. These oscillations tend to disappear as $W$ is increased, since the disorder mixes the wave-functions belonging to different sectors. The height of the plateau slightly decreases at first and then starts to increase again for $W > W_\star$, indicating that in the crossover regime the addition of quenched disorder {\it{enhances correlations between eigenstates}}.
As explained above, the plateau observed at small energy separation corresponds  to the energy scale (the so-called Altshuler-Shklovskii energy, defined as the Thouless energy at the scale of the lattice spacing) within which eigenstates exhibit GOE-like correlations. The plots thus show that in fact at small disorder the width of the energy window within which spectral correlations are GOE-like is much larger than the mean level spacing (vertical dashed lines). This result is in full agreement with the recent findings of~\cite{grimmroemer}, which indicate that the statistics of the gaps between states belonging to the same symmetry sector are described by the GOE ensemble. As one considers states closer to the spectral edges (right panel), the Altshuler-Shklovskii energy gets larger and the height of the plateau gets closer to the GOE universal value $K_2 (\omega) = 1$, indicating that in the small disorder limit the eigenstates close in energy near the edges are less correlated compared with eigenstates close in energy in the bulk of the band. For energy separation larger than the Altshuler-Shklovskii energy $K_2(\omega)$ shows a power law decay, $K_2(\omega) \sim \omega^{- \mu}$, with a disorder-dependent exponent which increases as $W$ is increased.  As the disorder is increased above $W_\star$ the plateau shrinks. Eventually at strong enough disorder the plateau disappears (i.e. it becomes smaller than the mean level spacing), and the standard localized behavior is recovered.

\subsection{Different wave-functions, different positions}

\begin{figure}
\includegraphics[width=0.431\textwidth]{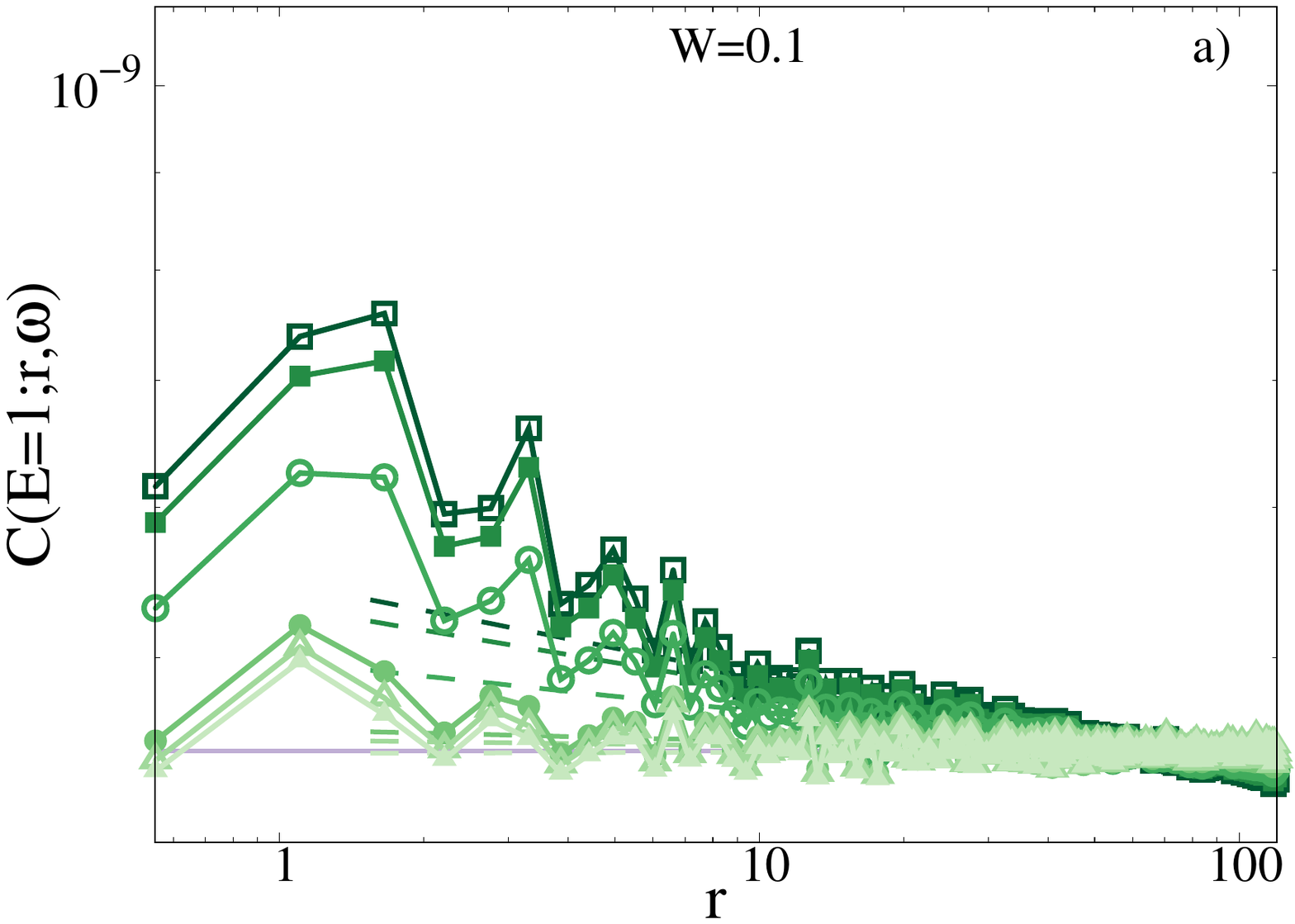} \hspace{0.1cm} \includegraphics[width=0.431\textwidth]{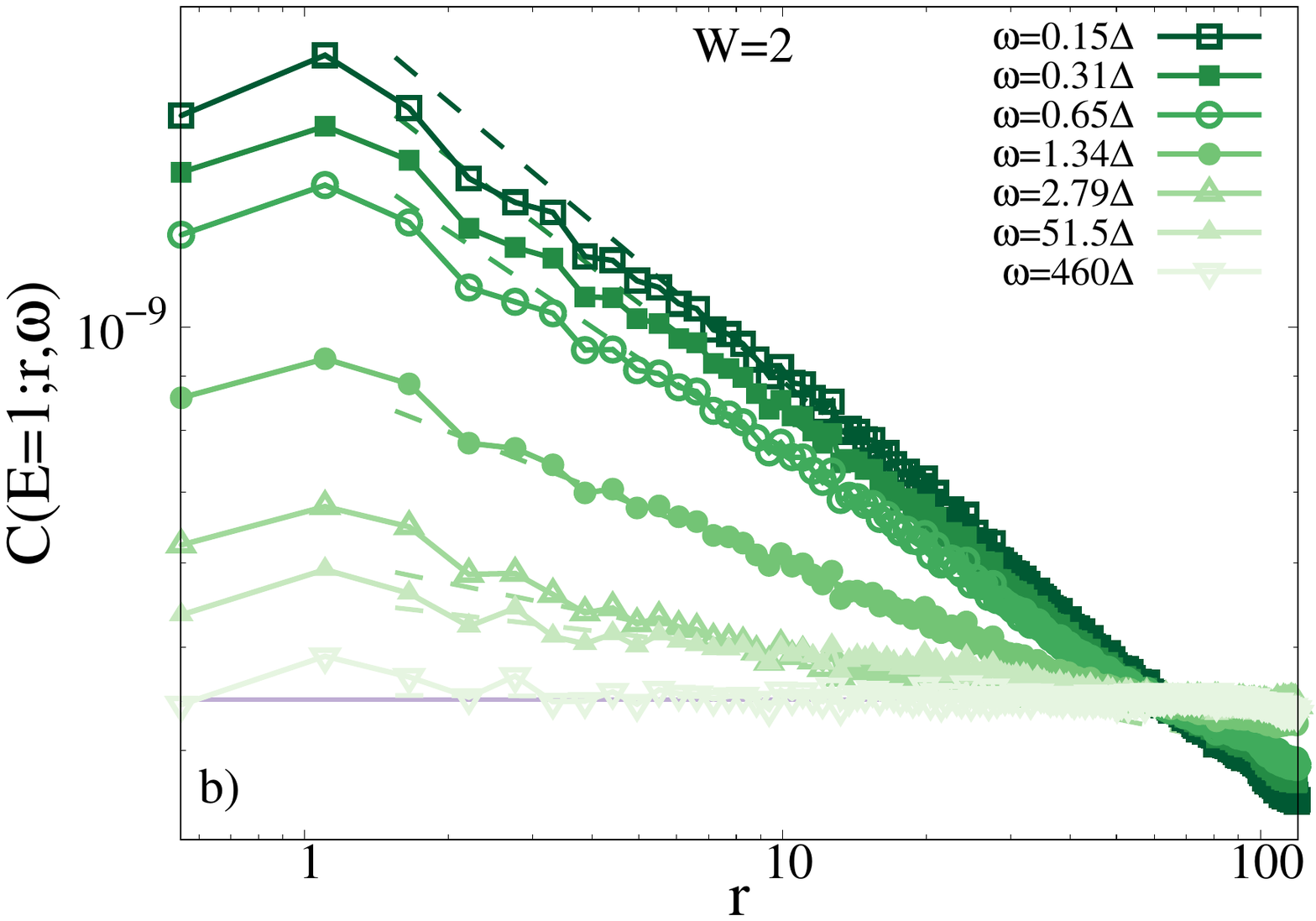} 

\hspace{0.1cm}

\includegraphics[width=0.431\textwidth]{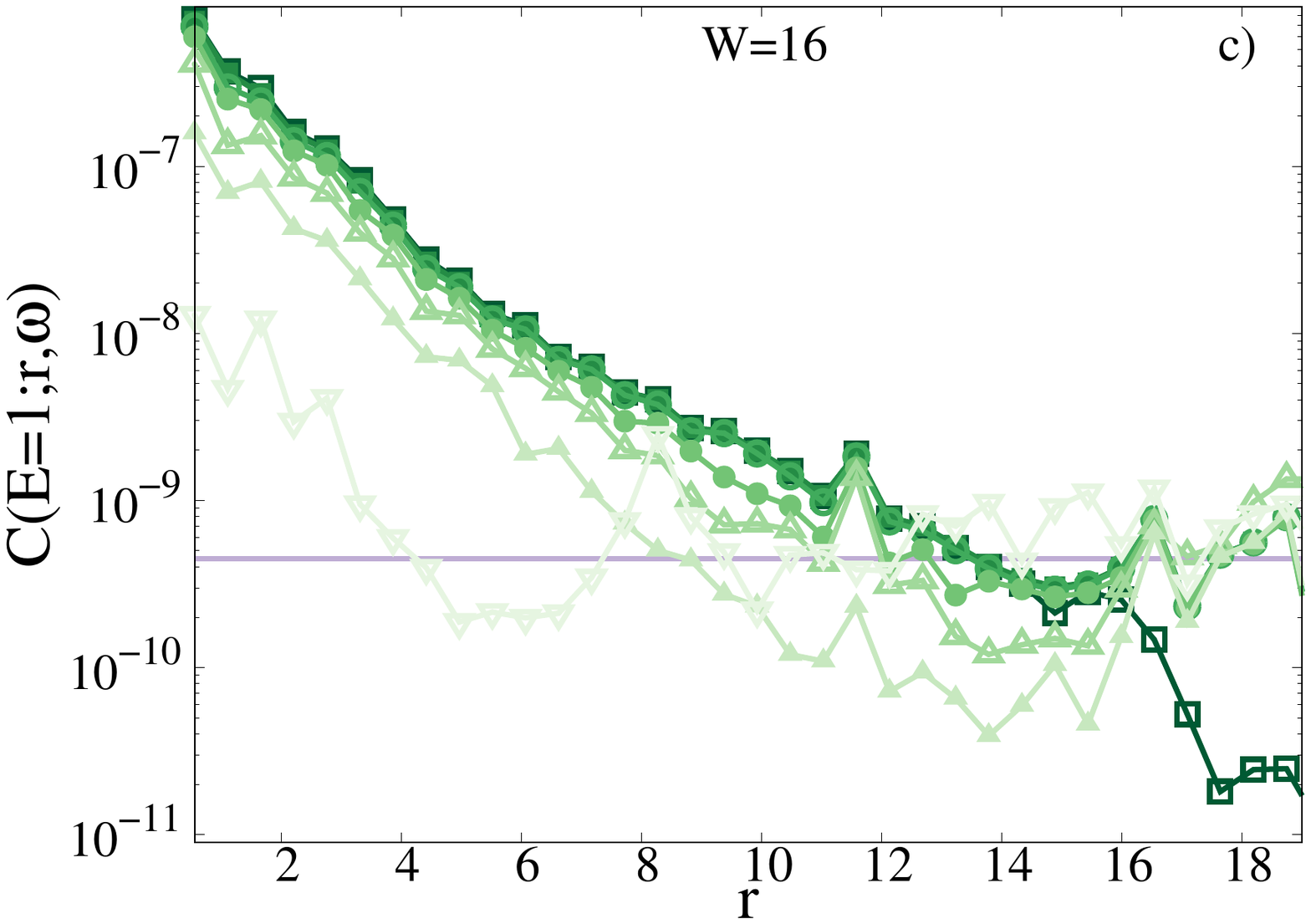} \hspace{0.15cm} \includegraphics[width=0.424\textwidth]{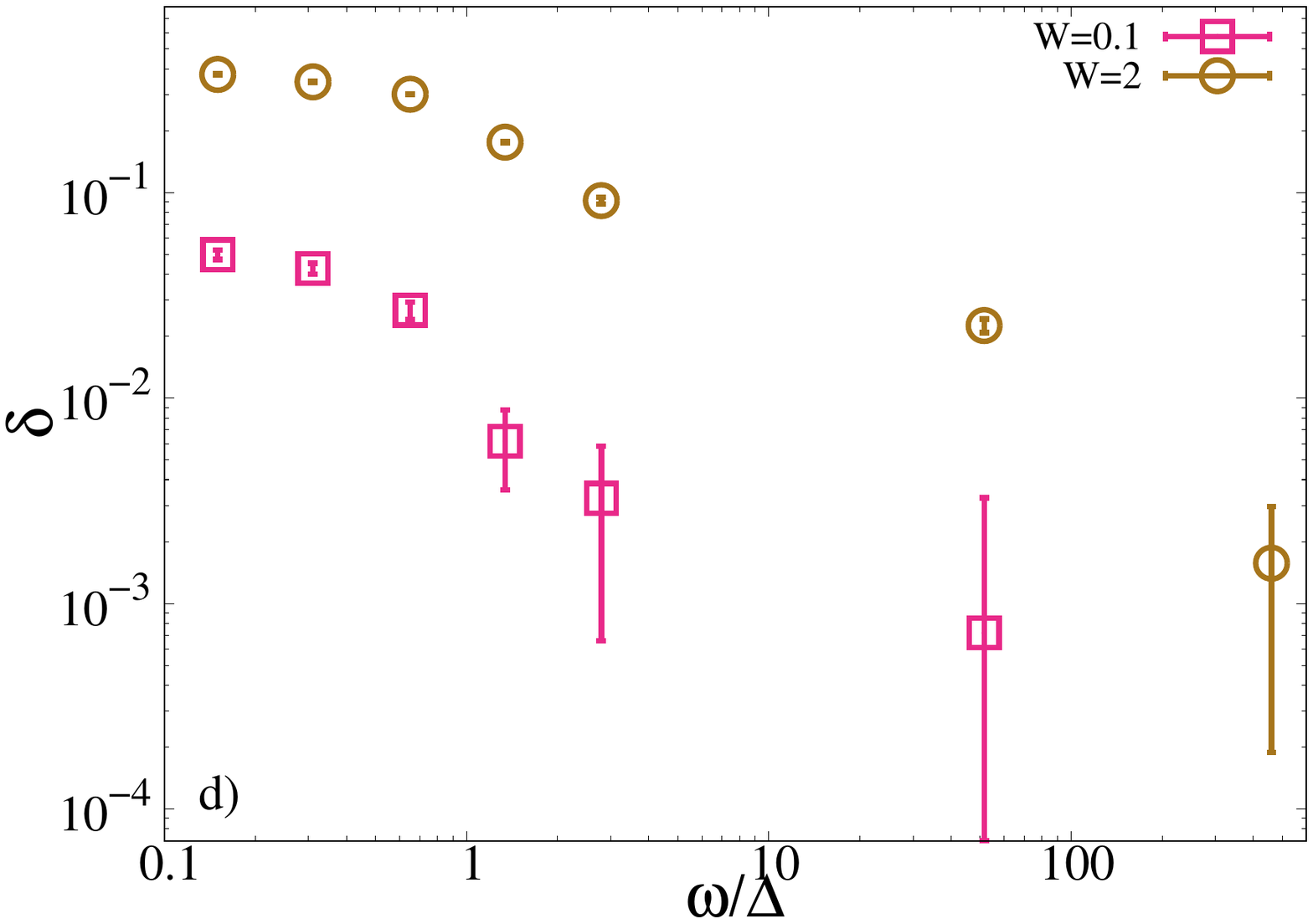} 
\caption{Correlation functions $C(E=1;r,\omega)$ between eigenstates of energy $E \pm \omega/2$ as a function of the distance $r = |r_1 - r_2|$ for $W = 0.1$ (a), $W=2$ (b), and $W=16$ (c), and for $N=47321$. Different colors corresponds to different energy separations $\omega$, measured in units of the mean level spacing $\Delta = 1/(N \rho(E=1))$ (where the average DOS $\rho(E=1)$ is measured at $W=0$). In a) and b) a  log-log scale is used  to highlight the effective power law regime in which the correlations decay as $C(E=1;r,\omega) \sim r^{-\delta (\omega)}$ (dashed lines), while c) is in a lin-log scale to highlight the exponential decay. Panel d) shows the evolution of the effective exponent $\delta(\omega)$ as a function of the energy separation (measured in units of the mean level spacing) for $W=0.1$ and $W=2$.
\label{fig:corr-energy}}
\end{figure}

Having characterized and interpreted the behaviors of $G(r;E)$ and $K_2(\omega;E)$, we now investigate the spatial correlations of the coefficients of different eigenstates.  Fig.~\ref{fig:corr-energy} shows data for a typical case corresponding to $E=1$ (similar results are observed for other band energies), and for three values of the disorder. The computations have been done for the approximant of size $N=47321$. Panels a and b  are log-log plots, so as to show the power law regime (dashed lines) more clearly, while the panel c is a linear-log plot so as to show the exponential behavior. Panel a corresponds to $W = 0.1 \approx W_\star$, where the eigenstates are the most extended. Panel b corresponds to $W=2$, in the crossover regime. Panel c corresponds instead  to $W=16$, in the strongly localized regime.  For $W=0.1$ and $W=2$ we observe that $C(E=1;r,\omega)$ decays as a power law at small enough energy separation, $C(E=1;r,\omega) \sim r^{-\delta (\omega)}$, with an exponent which depends strongly on the energy separation (dashed line). The figures show that the correlations become weaker and weaker as the energy separation is increased, in agreement with the fact that, as discussed above, $K_2 (\omega;E)$ is a decreasing function of $\omega$ (see Fig.~\ref{fig:K2}). Panel d shows the evolution of the effective exponent $\delta$ -- extracted by fitting the data of the top panels with power laws -- as a function of the energy separation $\omega$ (measured in units of the mean level spacing $\Delta=1/(N \rho(E))$). When the energy separation becomes much larger than $\Delta$, the effective exponent $\delta(\omega)$ becomes very small  and $C(r,\omega;E=1)$ approaches $1/N^2$, corresponding to uncorrelated wave-functions. Note that at $W=2$ the correlation function is much larger than at $W=0.1$, indicating once again that in the crossover regime the quenched disorder has the effect of enhancing the correlations among eigenstates. At strong disorder ($W=16$, panel c), deep into the strong Anderson localized regime, $C(r,\omega;E=1)$ decays exponentially with the localization length $\xi_{\rm loc}$ (for this value of W=16, $\xi_{\rm loc} \approx 5$ in our units).  Again, as noted in the case of the $K_2$ function, the correlations between different eigenfunctions becomes weaker as their energy separation is increased.

To conclude this section, we have seen that the correlations between states in the disordered 2D quasicrystal can be understood in terms of a qualitative picture as follows. For weak disorder and for a typical band energy, two states having a small energy difference $\omega \lesssim \Delta$ remain correlated at short distance, after which the correlations fall off as a power law. Increasing the disorder in the crossover regime increases the correlations, while also widening the range of the power law decay. Upon increasing the disorder strength beyond the crossover value, the  correlations fall exponentially rapidly. 

\section{Dynamics} \label{sec:dyn}

The evolution of states under disorder and their non-monotonic behavior at the characteristic crossover scale $W_\star$ can be expected to have a strong impact on relaxation, transport, and diffusivity. In particular, it is natural to expect that for $W$ close to $W_\star$, where the eigenstates appear to be more delocalized compared to the pure limit due to the effect of the random perturbation, transport and relaxation might be enhanced and accelerated. To check these ideas we run a few dynamical experiments in which a wave-function $| \phi_0 \rangle$ localized in a small spatial region spreads under the unitary evolution. If one chooses the initial amplitude to be localized on a single point this corresponds to a wave packet of energy $\mathcal{E}=\langle H\rangle \approx 0$ in the weak disorder regime ($\mathcal{E}=0$ in the pure limit). However, in this case the presence of a macroscopic number of confined states can be expected to affect dynamical properties. In the following we will instead consider initial conditions where the wavepacket is localized on``star clusters'', corresponding to non-zero energies. This is done by identifying all nodes of the lattice having $k$ neighbors ($3\leq k \leq 8$ for the AB lattice, see Fig.~\ref{fig:tiling}) and pick one of those nodes at random. The initial state is one in which the wave-function is localized on this node (index $i_0$) and its $k$ neighbors $i_\ell$ as follows:

\begin{equation} \label{eq:star}
| \phi_0 \rangle = \mathcal{C} \left( \cos \theta | i_0 \rangle \pm \sum_{\ell=1}^k \sin \theta | i_\ell \rangle \right) \, ,
\end{equation}
where the amplitudes are controlled by the parameter $\theta\in [0,\pi/2]$. $\mathcal{C}=1/\sqrt{1-(k-1)\sin^2\theta}$ is the normalization constant. Decomposing the Hamiltonian~\eqref{eq:hamiltonian} as $H = H_0 + {\cal D}$, where $H_0$ is the kinetic term ($t$ times the adjacency matrix of the AB tiling) and ${\cal D}$ the diagonal random potential, the energy of the initial star cluster state can be computed as a function of $k$ and $\theta$ as:

\[
\begin{aligned}
\label{eq:initialcon}
\mathcal{E} &= \langle \phi_0 | H | \phi_0 \rangle = \mathcal{E}_0 + \mathcal{E}_D \, , \\
\mathcal{E}_0 & = \langle \phi_0 | H_0 | \phi_0 \rangle =  \mathcal{C}^2 t k \sin 2\theta  \, , \\
\mathcal{E}_D & = \langle \phi_0 | {\cal D} | \phi_0 \rangle = \mathcal{C}^2 \left( \cos^2 \theta \, \epsilon_{i_0} + \sin^2 \theta  \sum_{\ell=1}^k \epsilon_{i_\ell} \right) \, .
\end{aligned}
\]
$\mathcal{E}_D$ is a random number of average zero and variance of order $(k+1)W^2$. The wavepacket energy can be varied by varying $k$ and/or $\theta$.
Note that Ref.~\cite{benza92} considers initial wave packets localized on a single site of coordination number $k$, which correspond to setting $\theta=0$ in the equation above. 

The time evolution of the wave-function at time $t$, $| \phi (t) \rangle$, can be written in terms of the eigenvalues $E_a$ and the eigenfunctions $| \psi_a \rangle$ of the Hamiltonian:

\[
| \phi (t) \rangle = \mathcal{C}\sum_{a=1}^N \left(\cos \theta \psi_a (i_0) \pm \sum_{\ell=1}^k \sin\theta \psi_a (i_\ell)\right) \, e^{-i t E_a / \hbar} |\psi_a \rangle \, .
\]
 Distances on the AB tiling are measured in units of $a$, the edge length of tiles. (To simplify the notation we set $\hbar=1$ throughout.)

\subsection{Probability of return to the origin}
The first observable we focus on is the so-called {\it return probability}, defined as the probability that a particle starting from a star state~\eqref{eq:star} at $t=0$ is found on the central node $i_0$ after time $t$. This is a measure of how fast the wave-packet spreads on the lattice and  decorrelates from the initial condition:
\begin{equation} \label{eq:rp}
P(t) = \left \langle \left \vert \langle i_0 | \phi (t) \rangle \right \vert^2 \right \rangle \, ,
\end{equation}
where the large brackets $\left \langle..\right \rangle$ denote the average performed over several realizations of the random potential, and several realizations of the initial condition (i.e., several choices of the central site $i_0$ among all the nodes with $k$ neighbors). 
In the limit $t=0$, this definition of the return probability gives $P(t=0) = \mathcal{C}^2\cos^2 \theta < 1$, while in the $t \to \infty$ limit $P(t)$ approaches a plateau value closely related to the IPR.

\begin{figure}

\includegraphics[width=0.431\textwidth]{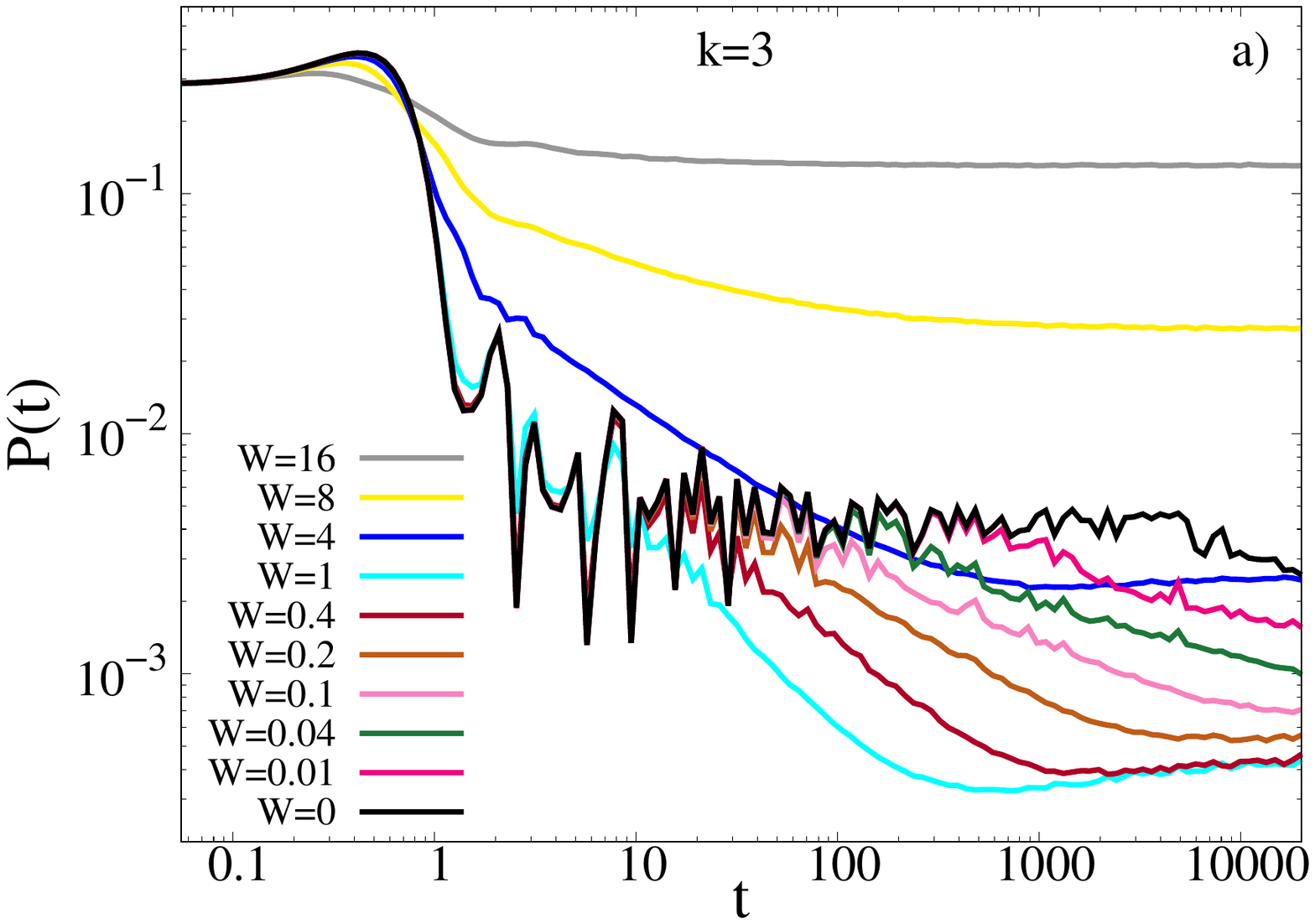} \hspace{0.1cm} \includegraphics[width=0.431\textwidth]{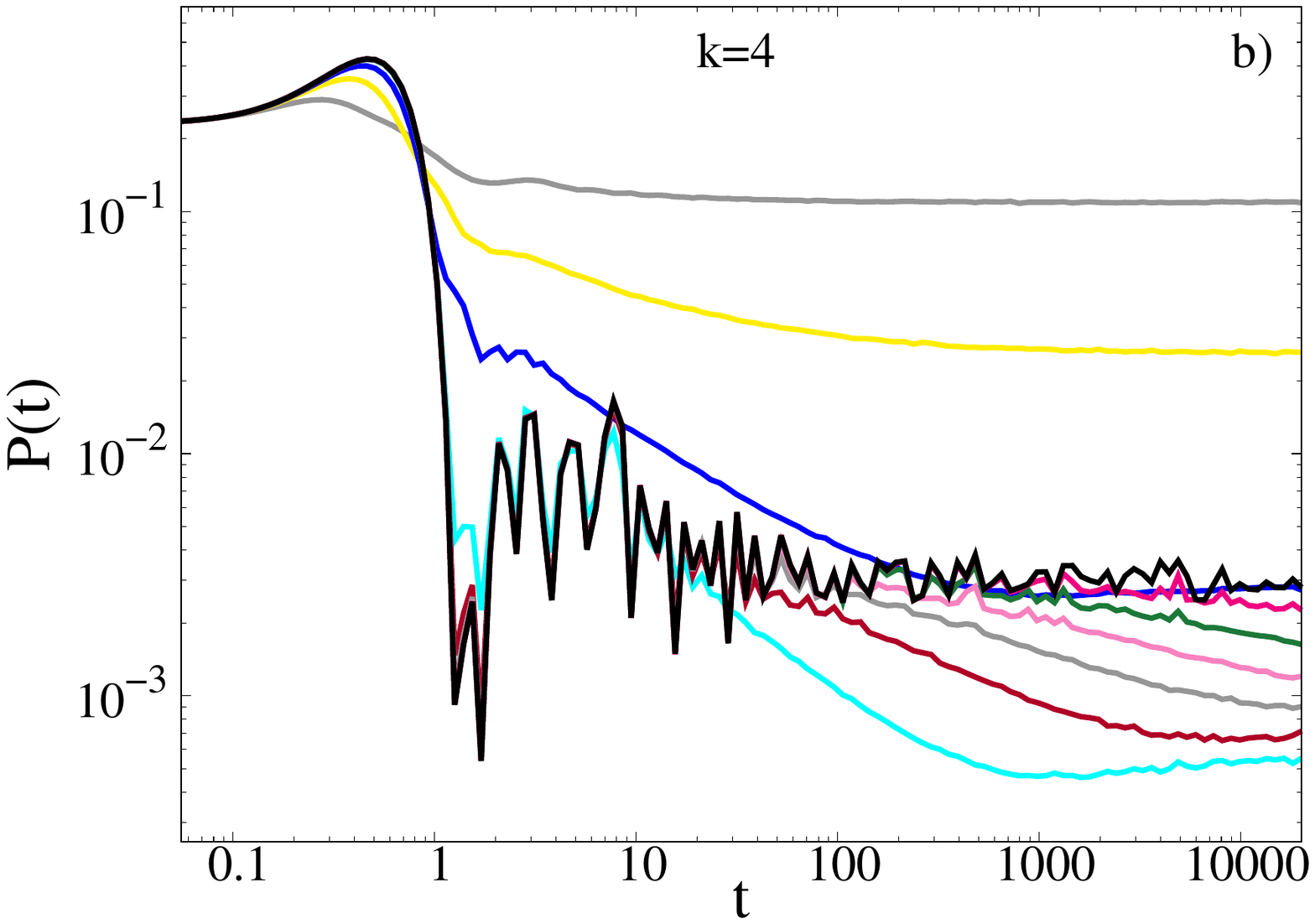}

\vspace{0.2cm}

\includegraphics[width=0.431\textwidth]{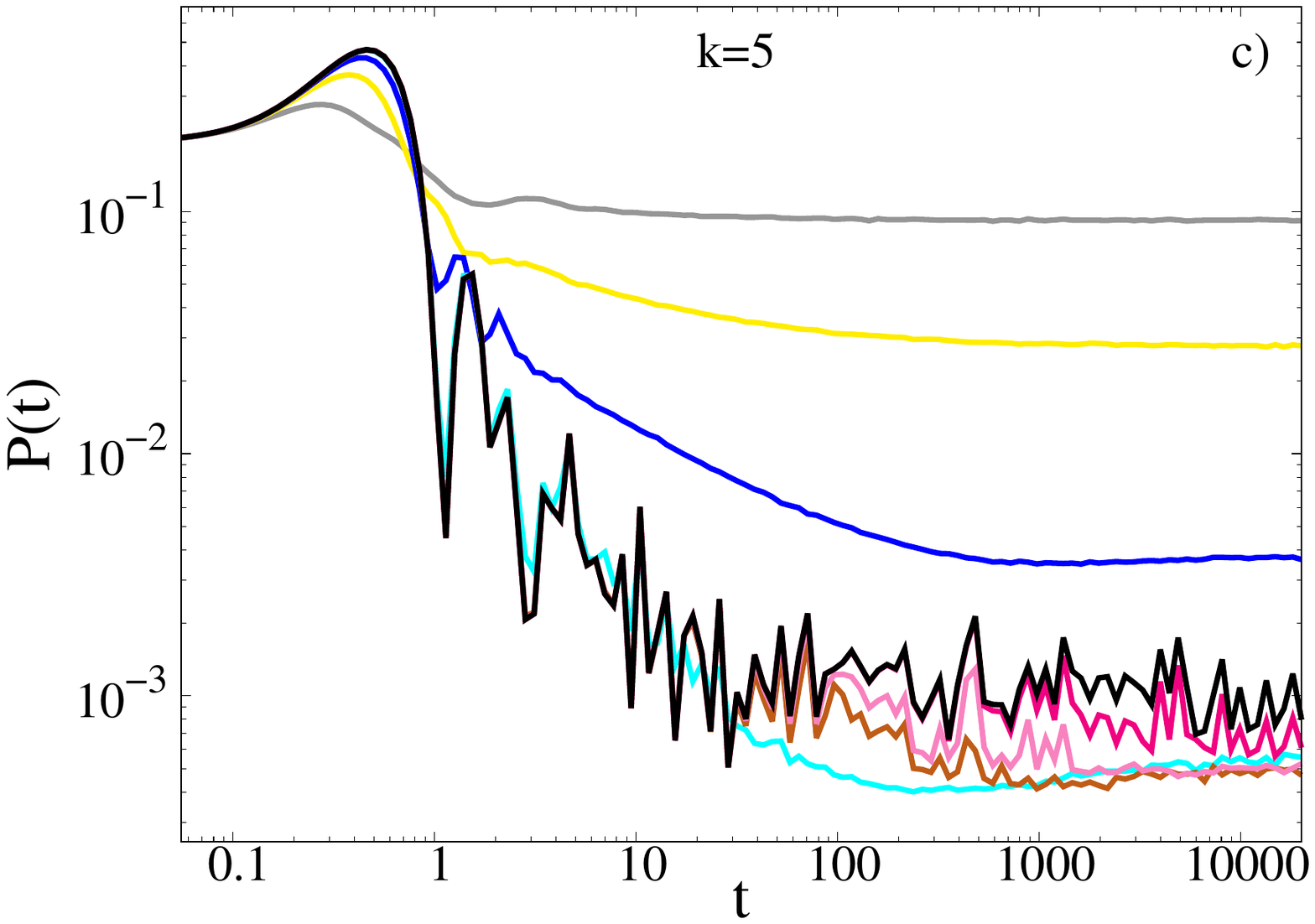} \hspace{0.1cm} \includegraphics[width=0.431\textwidth]{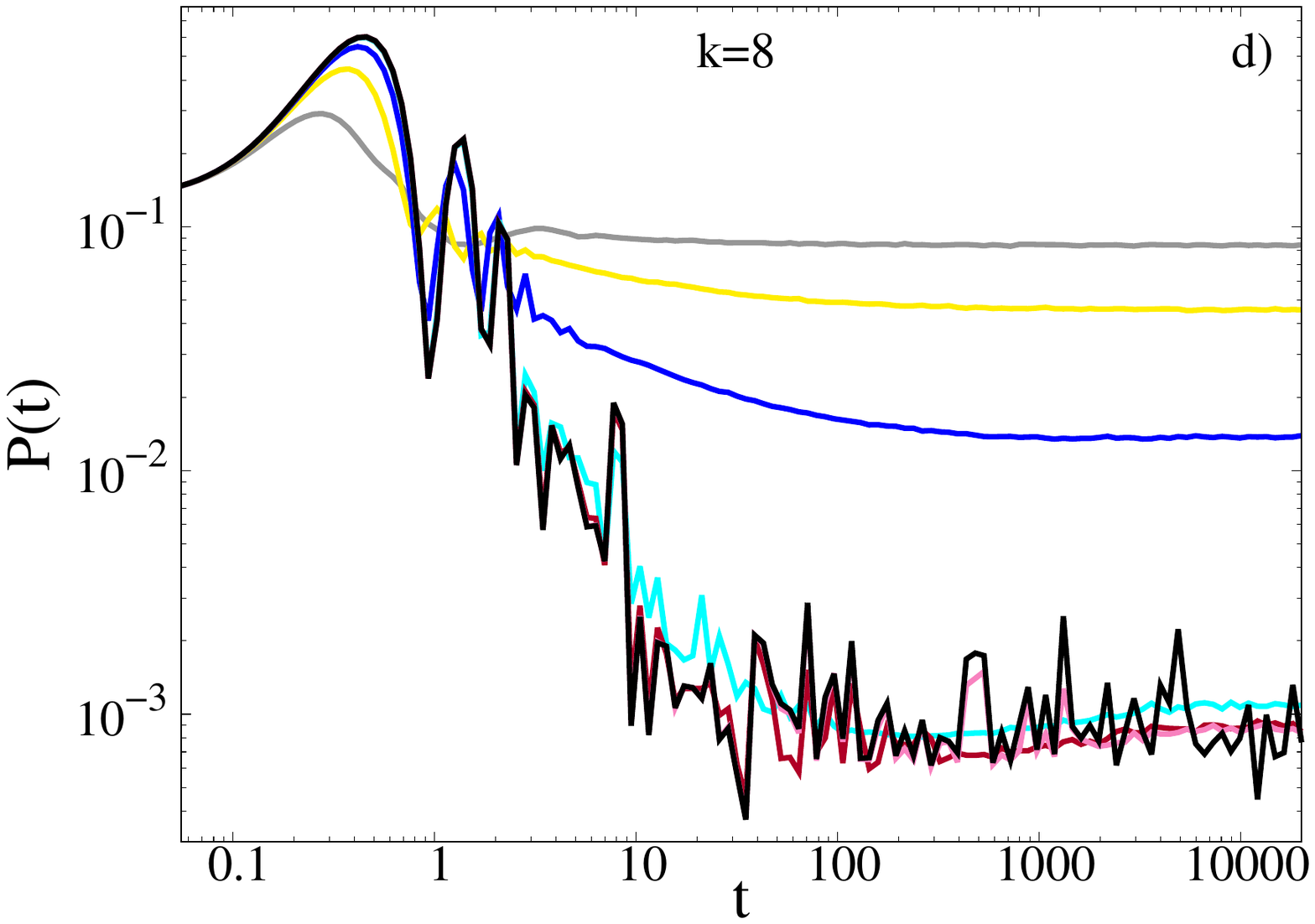}

\caption{Return probability~\eqref{eq:rp} as a function of time (in units of $\hbar$) for several values of the disorder and for $N=8119$. The four panels correspond to different choices of the initial conditions corresponding to star states described by Eq.~\eqref{eq:star} with $E=1.564$ (a), $E=1.684$ (b), $E=1.765$ (c), and $E=1.903$ (d). 
\label{fig:rp}}
\end{figure}

The numerical results (for $N=8119$) are plotted in Figs.~\ref{fig:rp} for four different choices of the initial conditions corresponding to star states with $k=3,4,5,8$ neighbors given in Eq.~\eqref{eq:star} and for $\theta \approx 0.741$. 
Figs.~\ref{fig:rp}a) and b) for 
$k=3$ and  
$k=4$ show that for these energies, disorder leads to a marked {\it decrease} of the return probability at intermediate and long times, as long as $W \lesssim W_\star$. This implies that introducing randomness indeed has the effect of speeding up the relaxation dynamics and the spreading of the wave-packet. In particular one sees that adding disorder produces a regime at moderately large times, $t \in [10^2,10^4]$, in which $P(t)$ has an apparent power law decay as a function of time. Such power law regime starts at earlier times upon increasing the disorder. We recall that when the initial wave-packet is localized on one site, the return probability is the Fourier transform of the spectral correlation function $K_2(\omega)$, and both have a power law dependence, given by the exponent $\mu$ of Sec.~\ref{sec:K2}. For our star-cluster initial conditions Eq.\eqref{eq:star}, the return probability is obtained upon averaging over correlation functions. The resulting $P(t)$ still shows a power law behavior at intermediate times, as a consequence of the power laws at intermediate energy separation of $K_2 (\omega)$. However the exponent of the time decay of $P(t)$ is no longer simply related to that of $K_2$. If the disorder is further increased above $W_\star$ then the infinite time plateau value of $P(t)$ increases significantly due to localization and the particle gets stuck in a small region of space around the initial state. 

Fig.~\ref{fig:rp}c) and d) show that the above described disorder induced acceleration of dynamics becomes weaker for 
$k=5$ (c), and it disappears completely for 
$k=8$ (d). 

\subsection{Mean square displacement and diffusion exponent $\beta$}
Another frequently studied quantity associated with quantum transport is the mean square displacement. This is defined as the average square distance from the central node of the initial star state $i_0$ traveled by the particle between time $0$ and $t$:
\begin{equation} \label{eq:msd}
\langle r^2 (t)\rangle  = \left \langle \sum_{j=1}^N | r_{i_0} - r_j |^2 \left \vert \langle j | \phi(t) \rangle \right \vert^2 \right \rangle \, ,
\end{equation}
where again the average is performed over several independent realizations of the disorder and several initial star states, when the central node is picked at random among all $k$th order nodes of the lattice. If one defines the diffusion exponent $\beta$ by $ \langle r^2 (t)\rangle \propto t^{2\beta}$, it can be easily shown that $\beta=1$ for ballistic motion (as in periodic crystals), while $\beta=1/2$ for classical diffusion (as in metals in the weak localization regime). In a system with localized states, the mean square displacement ceases to increase at large $t$ and $\beta=0$. Numerous quasiperiodic and random tiling models have been studied in the literature, with values of $\beta$ found to lie between 0 and 1, indicating a wide range of behaviors from super- to sub-diffusive motion depending on the model parameters \cite{benza92,choy,tsunetsugu,triozon,zhongmoss,trambly17,roche,vidal03}.

\begin{figure}
\includegraphics[width=0.327\textwidth]{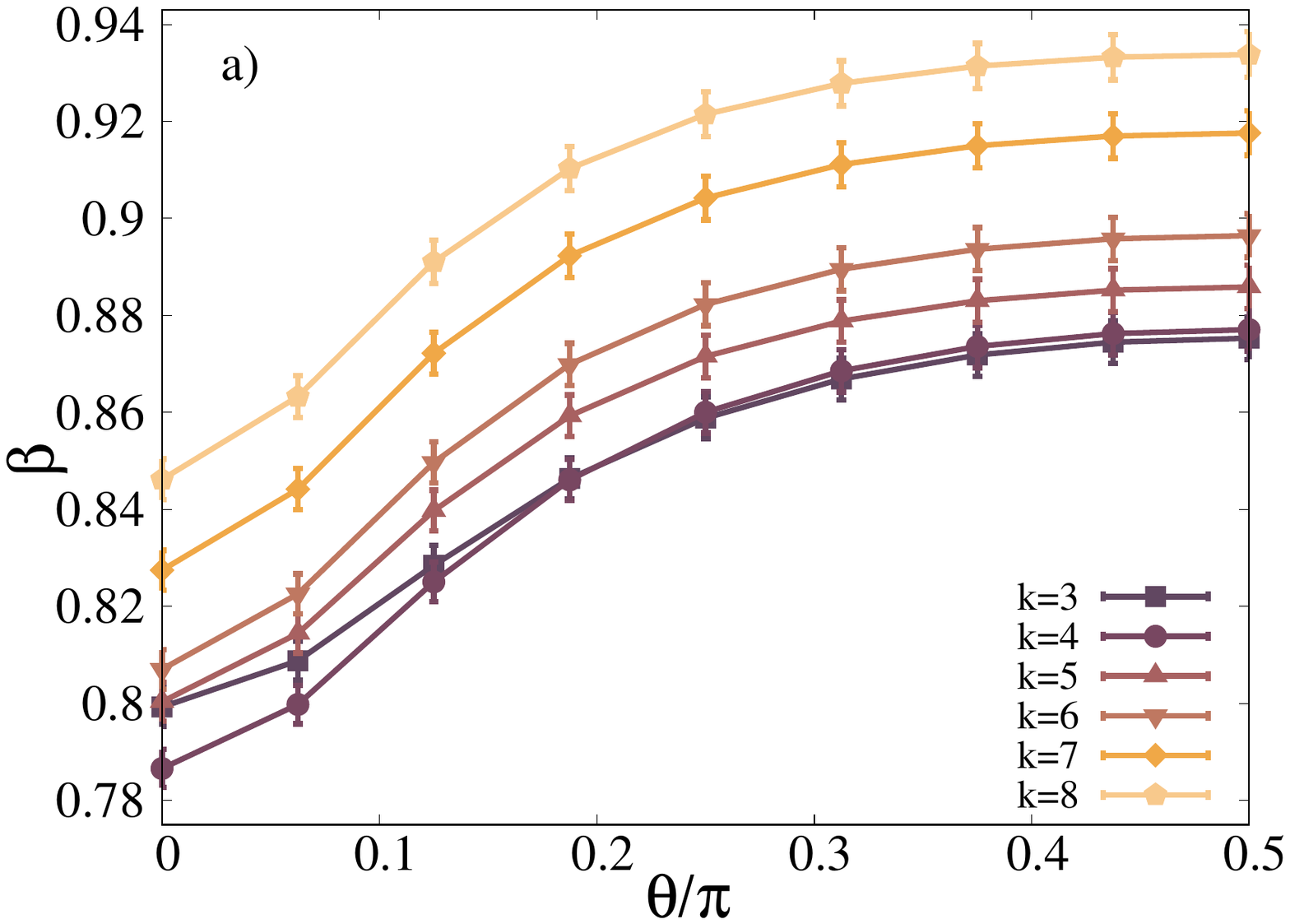} \hspace{-0.1cm} \includegraphics[width=0.325\textwidth]{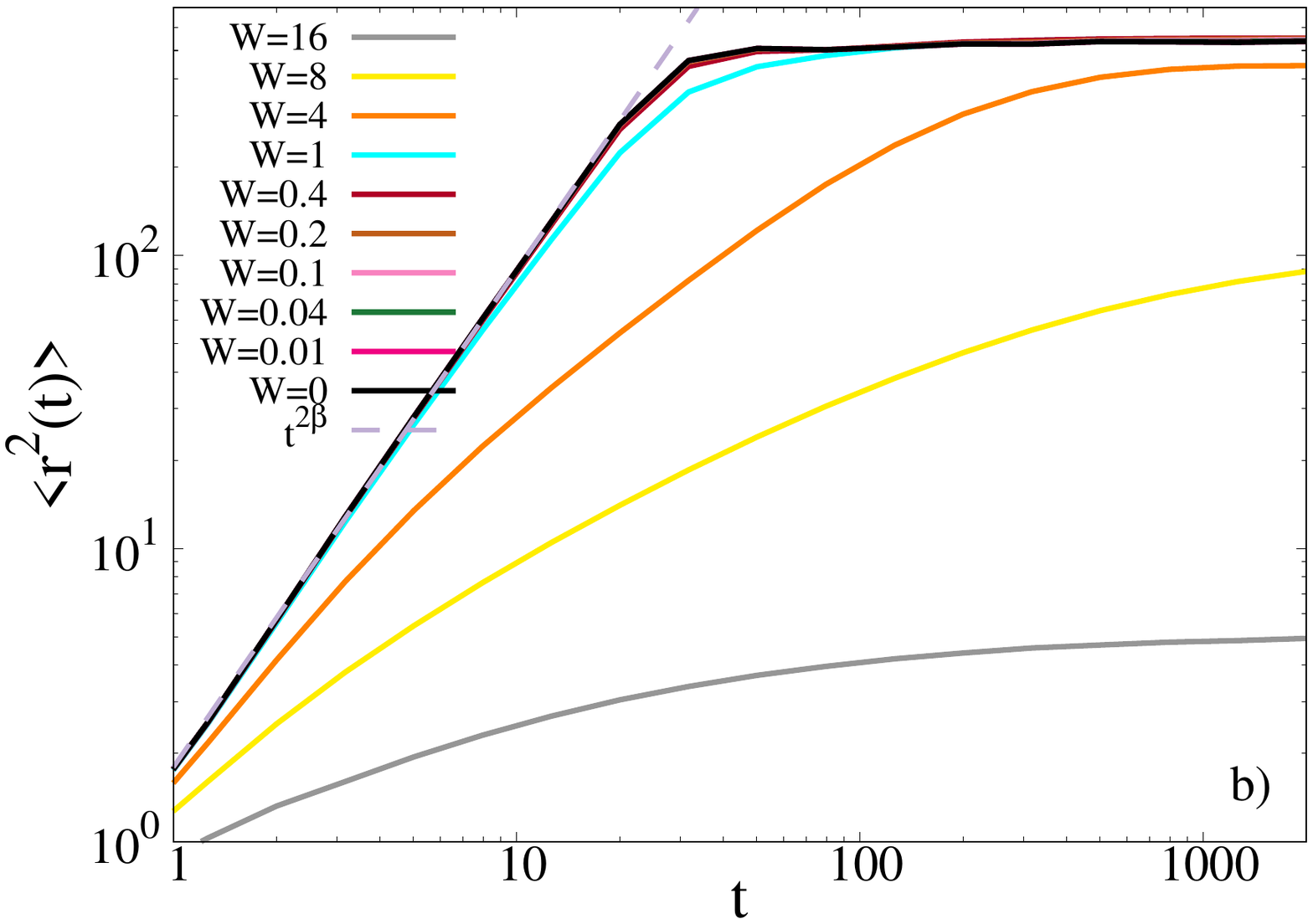} \hspace{-0.1cm}
\includegraphics[width=0.338\textwidth]{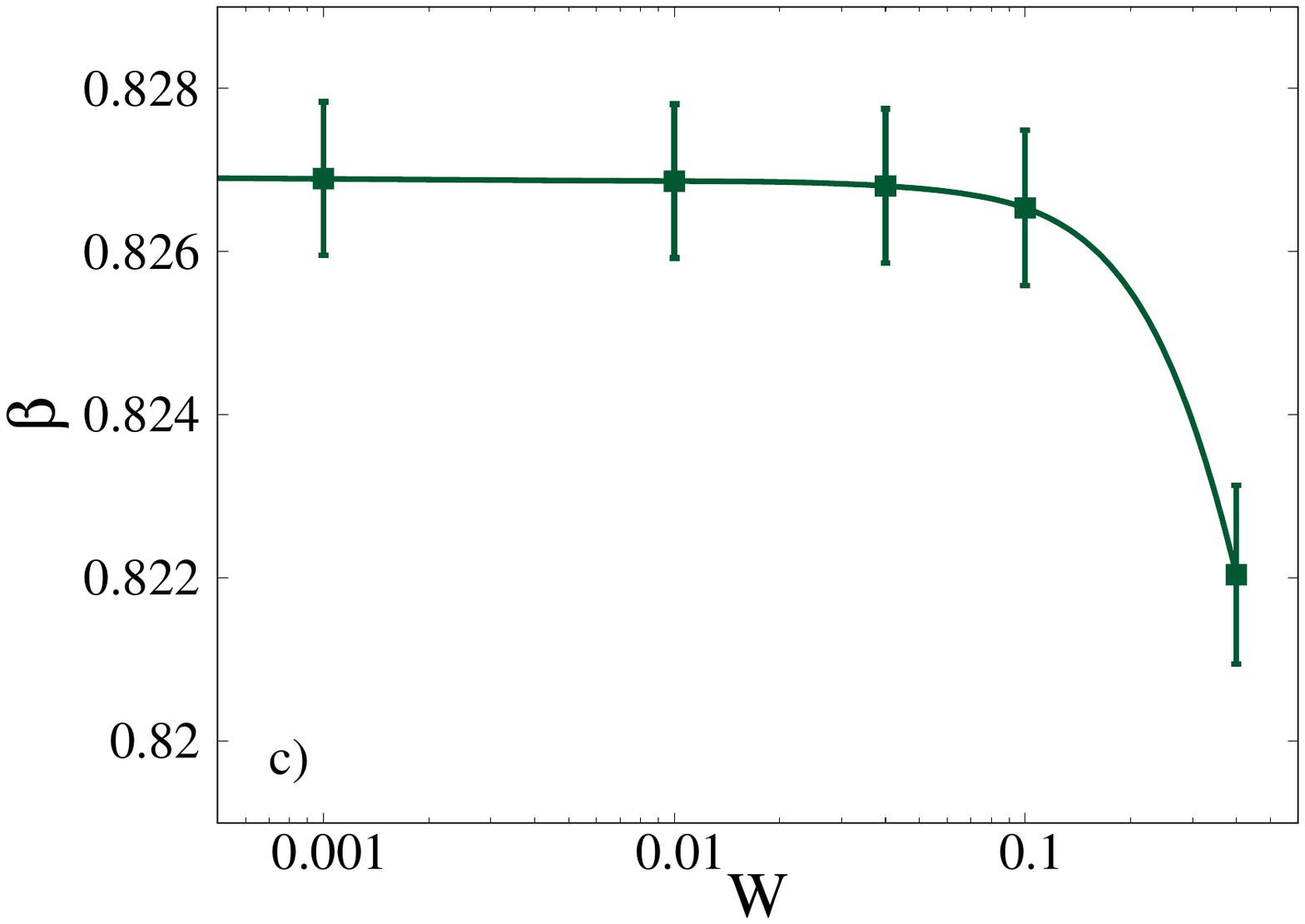} 
\caption{a) Diffusion exponent $\beta$ in the pure limit $W=0$ as a function of the parameter $\theta$ Eq.~\eqref{eq:star}, for all the values of the coordination number $k$ ($N=8119$). b) 
Mean square displacement as a function of time (in units of $\hbar$) for several values of the disorder, for $N=8119$, and for initial star states with $k=3$. The dashed straight line is a power law fit of the super-diffusive behavior as $\langle r^2 (t) \rangle \propto t^{2 \beta}$. c) Values of $\beta$ obtained in the weak disorder regime, $W \lesssim W_\star$, for $\theta = \pi/8$ and $k=4$ (similar results are obtained for other values of $k$ and $\theta$).}
\label{fig:msd}
\end{figure}

We start by studying how the diffusion exponent $\beta$ in the pure limit ($W=0$) depends on the local environment, and on the wave packet energy, governed by $\theta$. To this aim in Fig.~\ref{fig:msd}a) we plot the value of $\beta$ as a function of the angle $\theta$ and the coordination number $k$ (for system size $N=8119$). We find that in all cases $1/2<\beta<1$, corresponding to a super-diffusive behavior. This plot shows that $\beta$ increases with both $\theta$ (in the interval $[0,\pi/2]$) and $k$ (except when going from $k=3$ to $k=4$). The exponent is bigger for larger coordination number, as there are more channels to diffuse outwards. The exponent is smallest for small $\theta$, when the wave packet energy is close to zero, and the dynamics is governed by the more localized states at the band center. Our results agree fairly well with the early results of Ref.~\cite{benza92}, which correspond to the $\theta=0$ limit. However there are differences even in this limit. We find, for example, that for $\theta=0$ the $\beta$ values for $k=4$ and $k=5$ coincide. In contrast, Passaro et al found $\beta$ values were the same for $k=5$ and $k=6$. Fig.\ref{fig:msd}a) shows that as the wavepacket energy increases, clusters continue to have different diffusion coefficients, with the exception of $k=3$ and 4 when the curves of $\beta$ seem to converge as $\theta$ is increased. 

We next investigate the effect of quenched on-site disorder on the mean square displacement and on the exponent $\beta$.
Fig.~\ref{fig:msd}b) shows $ r^2 (t) $  for initial cluster star states~\eqref{eq:initialcon} 
with $k=3$ and $\theta = 0.741$.   
These illustrate that for small disorder, $W \lesssim W_\star$, the super-diffusive behavior is maintained at short times with an essentially unchanged power law with respect to the pure case. Then,
for larger $t$ the particle hits the boundaries so that $ \langle r^2 (t)\rangle $ becomes of the order of $L^2$ and saturates. As the disorder is increased above $W_\star$, the short time diffusivity is clearly diminished, and $ \langle r^2 (t)\rangle$ saturates at large times to the square of the localization length that becomes significantly smaller than $L$. 

In contrast to the return probability, the value of the diffusion exponent $\beta$ obtained from fitting the power law regime of $\langle r^2 (t)\rangle $ does {\it not} exhibit any non-monotonic behavior in the weak disorder regime. On the contrary, $\beta$ is essentially insensitive to disorder up to $W_\star$ within our numerical accuracy (and possibly even decreases very weakly with $W$). For $W \ge W_\star$, finally, $\beta$ starts to decrease steeply and vanishes at strong disorder in the localized regime. This behavior is shown in Fig.~\ref{fig:msd}c), where we plot $\beta$ {\it vs} $W$ for $\theta=\pi/8$ and $k=4$ (similar results are obtained for other values of $k$ and $\theta$). 

The above result for $\beta$ at weak disorder is to be contrasted with previous works on the effects of phason flip disorder on dynamical spreading. In \cite{benza92}, it was found that phason disorder tends to speed up quantum diffusion, as seen by the fact that the globally averaged $\beta$ value was thereby increased from 0.78 to 0.81. This discrepancy between their results and ours seems to indicate that the nature of the disorder plays an important role -- that on-site disorder and phason-flip disorder may have different effects on the spreading of the wave packet and on quantum transport. Indeed an important difference between the two forms of disorder is that, in our model with energetic disorder, sites conserve their connectivity, and the structure remains invariant under an inflation (tile rescaling) transformation. In contrast, phason flips lead to modifications of connectivities of sites and the loss of self-similarity under inflation in a geometrically disordered tiling. 

\subsection{Exponents and inequalities}
In the 1D case, a number of bounds have been proposed for diffusion exponents in terms of the generalized dimensions of the spectrum, $D_q$, and the wave functions $D_2^\psi$. A number of these have been generalized to 2D models, such as quasilattices generated by products of chains \cite{yuan2000,thiem2013}. These bounds have not however been checked for the 2D systems such as Ammann-Beenker and Penrose. It is nonetheless interesting to compare our results for the AB tiling with the bounds proposed in the literature. In the case of the hopping model on the pure AB tiling, it is believed that the spectrum is continuous and we expect therefore that $D_q=1$. The same is almost certainly true of the weakly disordered tilings. The Guarneri inequality \cite{guarneri}  generalized to dimension $D>1$ reads $\beta>D_1/D=0.5$. Thus this inequality is certainly obeyed in weakly disordered tilings whatever the energy $E$, since $\beta\gtrsim 0.8$.  Another inequality, due to Ketzmerick et al \cite{ketzmerick97} namely $\beta \geq D_2/2D_2^\psi \approx 0.56$ is also seen to hold in the present model. 

\section{Hopping conductivity in quasicrystals} \label{sec:vrh}
\begin{figure}
\includegraphics[width=0.5\textwidth]{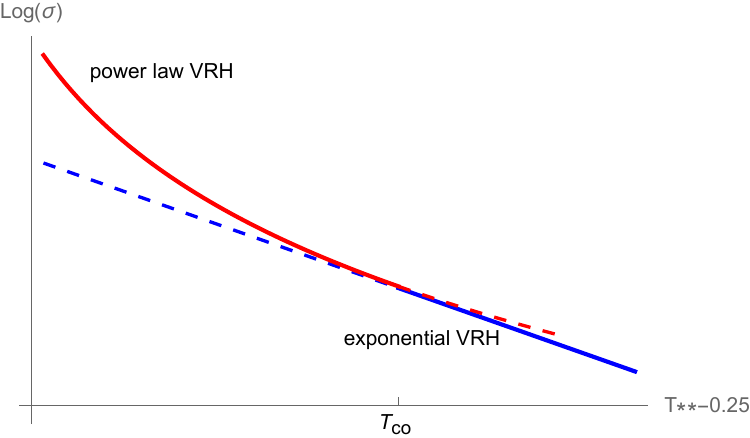} \hspace{-0.1cm} 
\caption{Sketch of the expected crossover of the temperature dependence of conductivity -- $\ln \sigma(T)$ is plotted versus $T^{-0.25}$ (arbitrary units), showing the transition from an intermediate $T$ power law VRH behavior (red curve) to a low $T$ exponential Mott VRH (blue curve). The crossover temperature $T_{co}$ corresponds to the condition $R_{mp}(T_{co}) \sim \xi_{\rm loc}$.
\label{fig.conduct}}
\end{figure}

In this final section we discuss some of the implications of our results for transport in three-dimensional systems. Many of our results can be expected to carry over to 3D although there are expected to be some important differences. In contrast with 2D, a genuine transition, rather than a crossover, is expected to occur in 3D systems for some $W=W_c$, beyond which all states will be localized. However, for finite 3D  systems (as when there are domain walls which introduce cutoff length for the quasiperiodic structure), one should see a similar evolution of states with disorder as we found in 2D. That is, for weak disorder the critical states of the pure 3D quasicrystal should initially delocalize, as is indeed observed by numerical studies \cite{olenev98,vekilov00}. These wave-functions then start to localize, and for a range of intermediate disorder strengths, based on our results, we expect that there will be a power law regime for wave function correlations. In this regime, 
transport of charge could occur via a variable range hopping mechanism for power law decaying wave-functions as was discussed first for disordered semiconductors \cite{mottkaveh}. The probability of hopping between two states near the Fermi level centered on two sites $r_1$ and $r_1+R$ and having an energy difference of $\omega$ can be estimated as
\begin{align}
\label{eq:proba}
     p_{hop} \propto R^{-\phi} e^{-\beta \omega} \, ,
\end{align}
where $\beta =(k_BT)^{-1}$ is the inverse temperature. The non-universal, disorder dependent exponent $\phi$ supposed positive, characterizes the spatial decay of the wave-functions near the Fermi energy, and is related to the exponent $\gamma$ of the function $G(r;E)$ by $\phi=\gamma-D$. Making the reasonable assumptions that i) the states are uniformly distributed in the sample volume $L^D$, and ii) assuming a locally flat DOS $\rho(E_F)$, which is true for intermediate disorder though not for pure systems, the mean energy difference between states is given by the condition
    $\omega \rho(E_F) R^{D} \sim 1$
where $D$ is the space dimension. The maximization of the hopping probability then leads to the most probable hopping distance as a function of temperature, $R_{mp}$ given by
\begin{align}
\label{eq:proba}
    R_{mp} \sim \left(\frac{D}{\rho(E_F) \, k_B T \phi}\right)^\frac{1}{D} \, ,
\end{align}
which decreases with the temperature. The resulting hopping conductivity is proportional to $\left(\frac{T}{T_0}\right)^{\phi/D}$, where $T_0 = (k_B \rho(E_F))^{-1}$  is a temperature scale determined by the DOS at the Fermi energy, which can be quite low due to the presence of a pseudogap. Eliminating $T_0$ in this expression, one concludes that the ratio of conductivities at two temperatures $T_1$ and $T_2$ within this temperature regime, is
\begin{align}
\label{eq:proba}
    \frac{\sigma_{VRH}(T_1)}{\sigma_{VRH}(T_2)} = \left(\frac{T_1}{T_2}\right)^{\phi/D} \, .
\end{align}
This power law VRH hopping conductivity $\sigma_{VRH}(T)$ can be expected in samples which are disordered enough that  $\gamma$ is large. If they are localized states, the necessary condition is that the localization length $\xi_{\rm loc}$ be larger than the most probable hopping distance $R_{mp}$ at the temperature $T$. However, when $T$ is lowered, such that $R_{mp}$ becomes larger than $\xi_{\rm loc}$, the standard exponential Mott VRH conductivity formula $\sigma^{Mott}_{VRH}\sim \exp[-(T_0/T)^{1/4}]$  should hold. We thus predict that, for disordered samples where $\xi_{\rm loc}$ is large, one should see a crossover between power law and exponential dependence when $T$ decreases, as sketched in Fig.~\ref{fig.conduct}. This type of crossover is in fact seen for the most highly resistive (quench cooled, polygrain) samples of AlPdRe measured in \cite{sarkar}. We should add that the power law hopping mechanism we have described here is $not$ expected to apply for perfect quasicrystals contrarily to some proposals in the literature \cite{janot}. The reason is clear: in a pure quasicrystal, the multifractal states are in general {\it{not}} localized around any particular position, and the notion of most probable hopping distance is undefined. 

\section{Discussion and conclusions} \label{sec:conclusions}
We have investigated disorder effects in quasicrystals by performing numerical studies in the 2D Ammann-Beenker tiling \cite{grunbaum}. We expect our findings to be valid at least qualitatively for other 2D quasiperiodic Hamiltonians with critical states, such as the Penrose rhombus tiling. Considering finite samples and starting from weak disorder, we have shown how the critical states of the pure model evolve as a function of the disorder strength. The main conclusion reached is that, for the majority of states, there is a non-monotonic evolution of spatial properties. They tend to delocalize initially, at weak disorder but then localize as expected for strong disorder, with localization lengths becoming smaller than the sample size.  
This interpretation is based on calculations of the generalized dimensions and on the multifractal spectrum $f(\alpha)$, for selected energies. We also computed 2-point correlation functions as a function of position and energy. These show that the multifractal character of wave-functions is preserved for weak out to moderate disorder in finite-size samples. For fixed system size, the crossover to strong localization occurs for a strength disorder $W_\star$, which depends on the energy, and on system size $N$. 

The initial tendency to delocalization of wave-functions due to disorder in quasicrystals is a robust phenomenon which is independent of the specific kind of disorder. This effect has been seen in previous studies of phason-disordered tilings~\cite{benza92,piechon95,schwabe,yamamoto,vekilov99,vekilov00}. It is an open question as to whether one could to reach the strong disorder regime solely by means of phason flips. It is possible that phason flip disorder, which preserves the chiral symmetry of the model, does not lead to localization, as has been observed in a different context in \cite{kawara}.

The above-described disorder dependence of the eigenstates' spectral properties and correlations has an impact on the quantum dynamics when starting from a wave-function initially localized in a small region of space. In particular the time dependence of the return probability of wavepackets $P(t)$ was seen to have different regimes. After an initial slow decay, $P(t)$ has a power law decay at intermediate times, and finally saturates at a value (basically the IPR) at very long times. The transition to power law behavior occurs at time $t_\star$, which moves to shorter times as disorder $W$ increases. As disorder is increased, the interval corresponding to power law decay shrinks, and the return probability saturates at higher values, showing the onset of strong localization. We have also studied the diffusion exponent $\beta$ for different initial conditions. In the pure tiling, the diffusion depends on the initial cluster size (is larger for sites with larger coordination number $k$) and increases as the wavepacket energy moves away from the band center. In contrast with the expectation based on earlier studies, we do not find a change of $\beta$ with weak added disorder. In the weak disorder regime, the exponent remains unchanged. An abrupt change of behavior is seen at the crossover disorder value $W_*$, beyond which $\beta$ drops steeply. For larger $W$ the strong localization regime is attained ($\beta=0$). These results are in contradiction with earlier studies on the effect of geometrical disorder due to phason flips on dynamical spreading of wave packets. Our study of dynamical spreading of wave packets seems once again to indicate that models with phason disorder are in a different universality class than the onsite disorder model. This question remains to be addressed in future studies, as well as questions of how the diffusion exponent depends on local environment and wavepacket energy.

Interestingly, the states around the pseudogap  behave differently from typical states within the band in that they do not show the initial tendency towards delocalization (upto our numerical accuracy) under weak disorder. Beyond a certain crossover disorder strength, these states evolve rapidly towards a strongly localized form. This is the case for the band edge states as well. In the disordered quasicrystal, as well as in Anderson disordered models, states at band edges are the first to show strong localization. This behavior of the pseudogap states could be pertinent for real quasicrystals: depending on the position of the Fermi level with respect to band edges, adding weak disorder could either enhance, or diminish transport. 
The eigenstate behaviors observed in the 2D model have also been observed in a 1D quasicrystal, the Fibonacci chain \cite{disorderFC1}. The main difference from 2D is that in 1D, non-monotonicity is seen only for a small subset of states. In 1D most of the states can be considered as band edge states -- quasiperiodicity results in gaps being opened throughout the spectrum -- and thus do not display non-monotonicity. 

As we said in the introduction, one of the motivations for this study is the problem of transport in quasicrystals. Our study provides some ideas for how transport depends on disorder in realistic 3D systems. We expect that, whereas in 1D and 2D, there is only a crossover from weak to strong disorder regimes for $W=W_\star$, for 3D, a "metal"-insulator transition should occur for some value $W=W_c$, where the term "metal" signifies that  there are no truly extended states. It is therefore more accurate to call this a "ergodic to nonergodic transition". For $W <W_c$, one should find a mobility edge separating exponentially localized states for $E>E_c(W)$ from critical states for $E<E_c(W)$.  In future work, it would be interesting to investigate properties of such a MIT by carrying out simulations for 3D quasiperiodic tilings. Based on our analysis of the 2D case, we propose a hopping mechanism which gives a prediction for the conductivity at low temperatures in disordered quasicrystals. This mechanism results in a crossover at some temperature  $T_{co}$, from an exponential Mott VRH conductivity to a power law conductivity. This type of crossover appears to be observed experimentally \cite{sarkar} for polygrain quasicrystal i-AlCuRe samples. To understand transport better, it will be necessary to carry out systematic measurements of the transport as a function of disorder strength. These are particularly challenging in this family of materials, which are difficult to synthesize. Samples are characterized by chemical and structural inhomogeneities, resulting in a number of controversies over the years \cite{fisher,dolinsek}.

Finally, the eigenstate correlation functions, and their evolution under disorder, have implications for numerous other electronic properties of tilings. The effects of disorder on the charge distributions studied in \cite{sakai2022}, on Kondo screening of impurities \cite{andrade}, valence fluctuations \cite{watanabe}, or properties of superconducting \cite{araujo,takemori} and magnetic phases \cite{wessel,koga} have not been addressed, with the exception of phason disorder effects in a Heisenberg antiferromagnet \cite{szallas}. It would be interesting to study how the charge distribution evolves with disorder. Last but not least, there are possibilities for realizing a 2D quasicrystal with cold atoms \cite{viebahn} and for studying interacting 2D models with quasiperiodicity \cite{coldatoms} indicates that such finely controlled and tunable systems could provide valuable new insights on the effects of disorder in quasicrystals.

\section{Acknowledgments}
We would like to acknowledge many useful discussions with  V. Dobrosavljevic. We would like to thank the referees for their helpful remarks, and for directing our attention to the work in ref.\cite{kawara}.

\end{document}